\documentclass{aa}  
\usepackage{graphicx}
\usepackage{txfonts}
\begin{document}

   \title{Chemical abundances of two extragalactic young massive clusters \thanks{Based on observations made with ESO telescopes at the La Silla Paranal Observatory under programme ID 084.B-0468(A)}}

   \author{Svea Hernandez
          \inst{1},
          S{\o}ren Larsen \inst{1},
          Scott Trager \inst{2},
         Paul Groot\inst{1},
          \and
   	Lex Kaper \inst{3}
          }
   \authorrunning{Hernandez et al.}

   \institute{Department of Astrophysics / IMAPP, Radboud University, PO Box 9010, 6500 GL Nijmegen, The Netherlands\\
          \email{shernandez@astro.ru.nl}    
          \and
          Kapteyn Astronomical Institute, University of Groningen, Postbus 800, NL-9700 AV Groningen, the Netherlands\
          \and
          Astronomical Institute Anton Pannekoek, Universiteit van Amsterdam, Postbus 94249, 1090 GE Amsterdam, The Netherlands\\
         }

   \date{accepted April 24, 2017}

 
  \abstract
   {}
   {We use integrated-light spectroscopic observations to measure metallicities
and chemical abundances for two extragalactic young massive star clusters (NGC1313-379 and NGC1705-1). The spectra were obtained with the X-Shooter spectrograph on the ESO Very Large Telescope.}
   {We compute synthetic integrated-light spectra, based on colour-magnitude diagrams for the brightest stars in the clusters from Hubble Space Telescope photometry and theoretical isochrones. Furthermore, we test the uncertainties arising from the use of Colour Magnitude  Diagram (CMD) +Isochrone method compared to an Isochrone-Only method. The abundances of the model spectra are iteratively adjusted until the best fit to the observations is obtained. In this work we mainly focus on the optical part of the spectra. }
   {We find metallicities of [Fe/H] = $-$0.84 $\pm$ 0.07 and [Fe/H] = $-$0.78  $\pm$ 0.10 for NGC1313-379 and NGC1705-1, respectively. We measure [$\alpha$/Fe]=$+$0.06 $\pm$ 0.11 for NGC1313-379 and a super-solar [$\alpha$/Fe]=$+$0.32 $\pm$ 0.12 for  NGC1705-1. The roughly solar  [$\alpha$/Fe] ratio in NGC1313-379 resembles those for young stellar populations in the Milky Way (MW) and the Magellanic Clouds, whereas the enhanced  [$\alpha$/Fe] ratio in NGC1705-1 is similar to that found for the cluster NGC1569-B by previous studies. Such super-solar [$\alpha$/Fe] ratios are also predicted by chemical evolution models that incorporate the bursty star formation histories of these dwarf galaxies. Furthermore, our $\alpha$-element abundances agree with abundance measurements from H II regions in both galaxies. In general we derive Fe-peak abundances similar to those observed in the MW and Large Magellanic Cloud (LMC) for both young massive clusters. For these elements, however, we recommend higher-resolution observations to improve the Fe-peak abundance measurements.}
{}

   \maketitle
%

\section{Introduction}\label{intro}
The study of stellar atmospheres and their composition can provide a detailed picture of the  chemical enrichment history of the host galaxy. Given that stellar atmospheres generally retain the same chemical composition as the gas reservoir out of which they formed, one can gain unparalleled knowledge of the gas chemistry of a galaxy throughout its formation history through the analysis of  stellar populations of different ages. Abundance ratios can be used as tracers of initial mass function (IMF) and star formation rate (SFR) and can provide a relative time scale for chemically evolving systems \citep{mcw97}. $\alpha$-elements (O, Ne, Mg, Si, S, Ar, Ca, and Ti) are primarily produced in core-collapse supernovae \citep{ww95}, which trace star formation in a galaxy, whereas Fe-peak elements (Sc, V, Cr, Mn, Fe, Co and Ni) are produced mainly in Type Ia SNe at later times \citep{tin79,matgre86}. In the case where there is a burst of star formation, Type II SNe appear first producing $\alpha$-enhanced ejecta. Later, as the Type Ia SNe appear, the ejecta become more Fe rich. Previous studies of the Milky Way (MW) have shown that [$\alpha$/Fe] ratios are particularly useful in identifying different stellar populations, their location and substructures \citep{ven04,pri05}. The halo and bulge stellar populations in the MW display an enhancement of $\alpha$-elements indicating that a relatively rapid star formation process took place \citep{wor98,mat03}. The populations identified via different abundance patterns (along with other information) have so far helped us develop and assemble a detailed nucleosynthetic history for our galaxy. \par

Prior to the advent of 8-10 m telescopes, extragalactic abundances of stars were limited to supergiants in the Magellanic Clouds \citep{wolf73,hill95,hill97,ven99}. These new telescopes and their instruments made studies of fainter stars (i.e. Red Giant Branch stars) outside of our own galaxy possible. The abundances obtained from stars in dwarf galaxy environments, for example, showed a clear difference when compared to the chemical evolution paths followed by any of the MW components \citep{shet98,shet01, bon00, tol03}. In general [$\alpha$/Fe] ratios at low [Fe/H] in dwarf galaxies resemble the ratios observed in the MW. However, as metallicity increases [$\alpha$/Fe] ratios in dwarf galaxies have been measured to evolve down to lower values than those seen in the MW for similar metallicities. It has been hypothesised that these low ratios might be caused by a sudden decrease of star formation, although this topic is still being debated \citep{tol09}. To establish a well understood pattern of star formation covering all mass ranges, detailed abundances beyond the MW and its nearest dwarf spheroidal neighbours are needed. \par

For external galaxies, especially beyond the Local Group, most of the chemical composition information comes from measurements of H II regions in star-forming galaxies \citep{sea71,rub94,lee04,sta05}. Such measurements, however, are limited to the present-day gas composition, and do not provide information on the past history of the galaxy in question. Furthermore, H II region studies do not usually provide constraints on [$\alpha$/Fe] ratios. Considering that individual stars are too faint for abundance analysis beyond the MW and nearby dwarf galaxies, we instead target star clusters, which are brighter and can therefore be observed at greater distances. In such analyses, it is usually assumed that the clusters are chemically homogeneous and consist of stars with a single age. \par 

With current telescopes we can obtain intermediate- to high-resolution spectra of unresolved star clusters out to $\sim$ 5-20 Mpc. The integrated-light spectra of most star clusters are broadened only by a few km s$^{-1}$ due to their internal velocity dispersions, which allows for higher resolutions ($\lambda/\Delta \lambda \sim$20,000-30,000) and makes the detection of weak ($\sim$15 m$\AA$) lines possible. Previous studies have developed techniques to extract detailed abundances from unresolved extragalactic globular clusters (GCs) at moderate S/N ($\sim$ 60) and high resolution ($\lambda/\Delta \lambda \sim$ 30,000). \citet{mc08} developed a technique to analyse high-resolution integrated-light spectra in which the abundances are measured using a combination of Hertzsprung-Russell diagrams (HRDs), model atmospheres and synthetic spectra. Their method employs similar procedures to those used in spectral analyses of single stars \citep{mc02, ber05} and has been tested and improved for integrated-light spectra of GCs as far as $\sim$ 780 kpc from the MW \citep{col09, col11, col12}. In addition to these detailed abundance studies of nearby GCs, \citet{col13} initiated a study of the chemical evolution and current composition of the old populations in NGC 5128. This investigation has obtained metallicities, ages and Ca abundances of 10 GCs in this external galaxy, 3.8 Mpc away. \par

\citet{lar12} (hereafter "L12") also created a general method to analyse integrated-light spectra that shares many of the basic concepts introduced by \citet{mc08}.  However, two of the advantages of the L12 technique are the modelling of broader wavelength coverage and solving simultaneously for a combination of individual element lines. This method then takes into account contributions from both weak and strong lines. One of the main motivations for the development of the L12 method was to create a more generalised analysis that can be used with data of different resolutions and S/N and is not exclusive to high-dispersion observations. L12 tested this new technique using high-resolution integrated-light observations of old stellar populations in the Fornax, WLM, and IKN dwarf spheroidal galaxies \citep{lar14} and determined metallicities and detailed abundance ratios for a number of light, $\alpha$, Fe-peak, and \textit{n}-capture elements. \par

Most of the detailed abundance studies outside of the MW have focused so far on the oldest stellar populations. Part of the reason for this is that GCs have extensively proven to be a useful tool for tracing both the early formation and the star formation throughout the history of the galaxy \citep{sea78, har91,wes04}.  In addition to tracing the history of galaxies, old stellar populations are relatively better understood than young ones: the HRDs of GCs are well understood and accounted for by models (with some exceptions concerning horizontal branch morphology), and the spectra of the types of stars found in GCs, for the most part, can be modelled using relatively standard techniques. While the results from these old populations in external galaxies are useful, to get a broader and fuller picture of the star formation history of these galaxies we need to extend this analysis to younger extragalactic populations. The combination of studies involving GCs and younger stellar populations allows us to observe the chemical evolution of galaxies through a much larger window in time.\par

Methods to study younger stellar populations, other than H II regions, include the analysis of evolved massive stars, blue \citep{kud08} and red \citep{dav10} supergiants.  Studies of blue supergiants (BSGs) have shown excellent agreement with H II region measurements \citep{kud12,kud14}. In addition, this method was successfully used in the first direct determination of a stellar metallicity in NGC 4258, a spiral galaxy approximately 8.0 Mpc away \citep{kud13}. Similar results have been found when studying red supergiants (RSGs). The galactic and extragalactic metallicities obtained through the use of RSGs agree quite well with measurements from BSGs of young star clusters \citep{gaz14} and are consistent with other studies of young stars within their respective galaxies \citep{dav15}. 

In the last couple of decades significant populations of Young Massive Clusters (YMCs) have been identified in several external galaxies with on-going star formation \citep{lar992, lar04}. YMCs are defined as young stellar clusters of ages < 100 Myr and stellar masses of > 10$^4$ $M_{\odot}$ \citep{por10}. Just like GCs, YMCs have increasingly been used as tracers of star formation, as they are considered the progenitors of GC populations \citep{zep93} and are found in regions with high star formation rates \citep{sch98,whi95}. \par

In an effort to continue exploring the chemical signatures of young stellar populations, \citet{lar06} and \citet{lar08} have measured metallicities and [$\alpha$/Fe] abundance ratios of single YMCs in NGC6946 and NGC1569. Both studies found super-solar [$\alpha$/Fe] ratios and metallicities of [Fe/H] $\sim-$0.45 dex for NGC6946-1147 and [Fe/H]$\sim-$0.63 dex for NGC1569-B.\par

\citet{col12} also studied several young clusters in the Large Magellanic Cloud (LMC). They measured metallicities, $\alpha$, Fe-peak and heavy element abundances for three young star clusters $<$ 500 Myr using integrated-light analysis. \citeauthor{col12} also studied three GCs and two intermediate-age clusters in the same galaxy. This star cluster sample allowed them to detect an evolution pattern in [$\alpha$/Fe] ratio with [Fe/H] and age, where younger star clusters were observed to have higher metallicities ($-$0.57$<$[Fe/H]$<$$+$0.03 dex). \par

Most of the progress in the study of these young stellar populations, including many of the methods mentioned above, has focused on the Near-Infrared (NIR) part of the spectrum. According to \citet{ori00}, the NIR stellar continuum at relatively young ages ( $\sim$10 Myr) is entirely dominated by the flux from RSGs. This has allowed a type of analysis where the spectrum of a YMC is approximated as that of a single RSG, and analysed as such. The scene changes when studying these populations at optical wavelengths, because the integrated-light spectrum of a YMC then cannot be approximated by that of a single RSG. In order to account for line blending in this wavelength regime, one needs to integrate full spectral synthesis techniques and account for every single evolutionary stage present in the cluster. With the aim to continue exploring this new territory we perform a detailed chemical study on two YMCs, primarily focusing on the optical part of the spectrum. \par

In this paper we present detailed analysis of two YMCs, NGC1313-379 and NGC1705-1. We study their chemical composition through the analysis of intermediate-resolution integrated-light observations. Using the same method and software developed by L12, we derive metallicities and detailed stellar abundances for both objects. For the first time we obtain abundances of $\alpha$ (Mg, Ca, Ti) and Fe-peak (Sc, Cr, Mn, Ni) elements of stellar populations in both NGC1313 and NGC1705. In Section \ref{obs} we provide a description of the target selection, science observations and data reduction. In Section \ref{ana} we explain the analysis method, including the atmospheric modelling, creation of the synthetic spectra, wavelength range and clean line selection, and provide a brief description of the main properties of the individual YMCs. Section \ref{results} describes the main results of this work, and in Section \ref{discussion} we discuss the implication of our measurements and compare them with chemical abundance studies for stars in the MW, M31 and the LMC. Finally in Section \ref{Conclusions} we list our concluding remarks.


   
\begin{figure*}
\centering
\includegraphics[width=7.5cm]{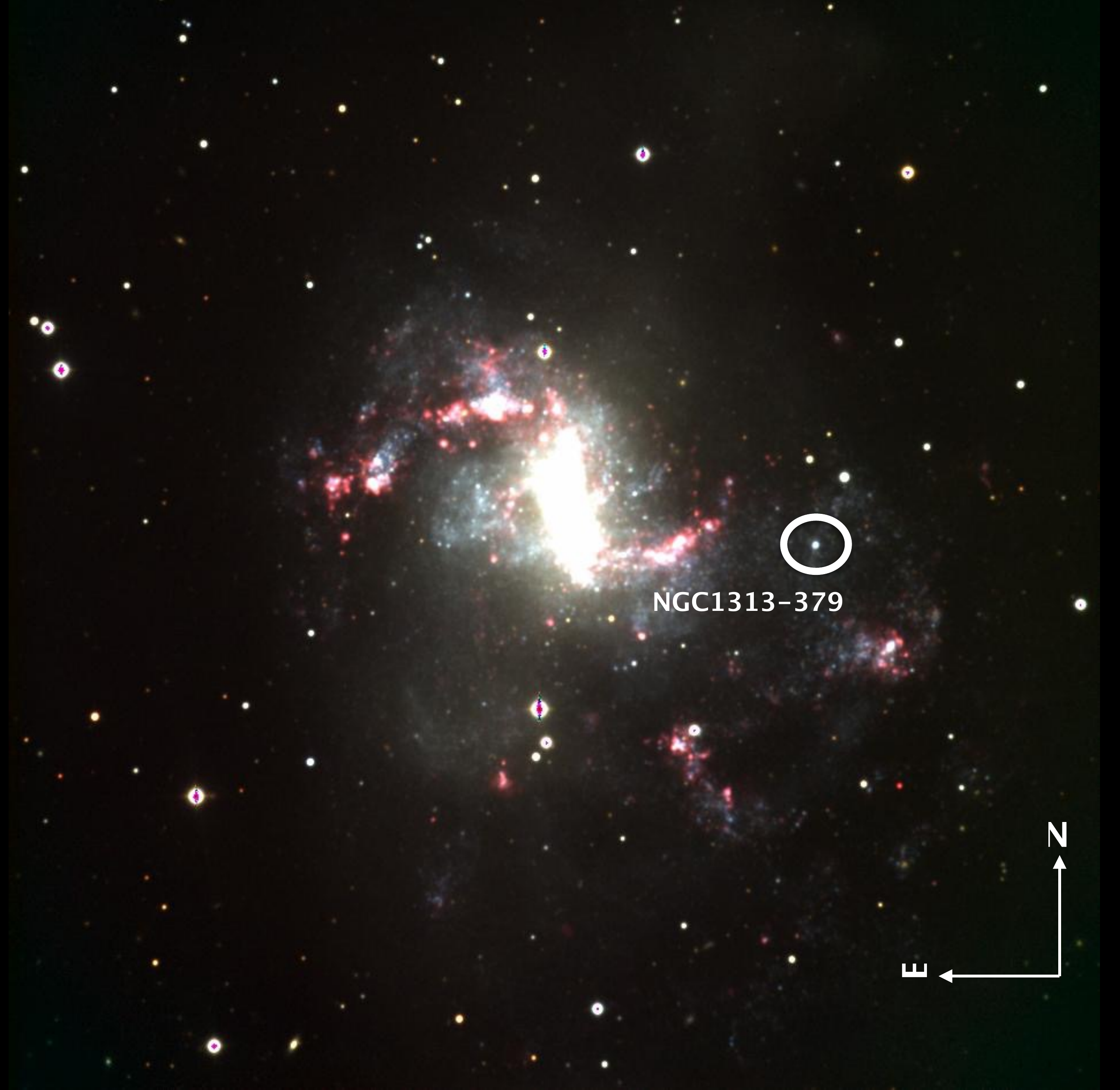} 
\includegraphics[width=6cm]{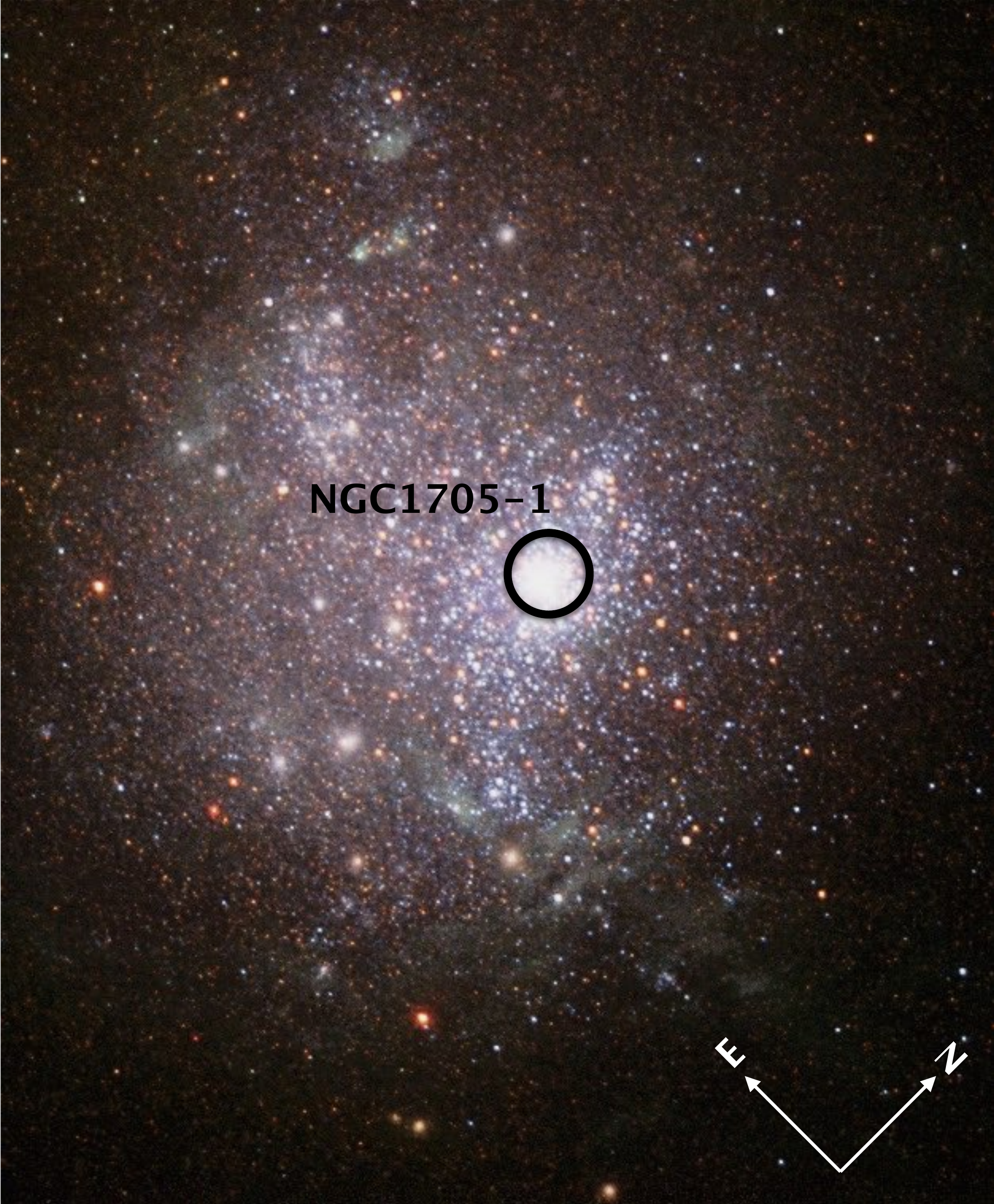}
\caption{Left panel: Colour-composite image observed with the ESO Danish 1.54-metre telescope located at La Silla, Chile. The white circle shows the location of the YMC NGC1313-379.  Image credit: \citet{lar99}; Right panel: Colour-composite image observed with the Wide Field Planetary Camera 2  on board the Hubble Space Telescope. YMC NGC1705-1 is shown in the black circle. Image Credit: NASA, ESA, and The Hubble Heritage Team (STScI/AURA).}
\label{fig:galaxies}
\end{figure*}

\section{Observations and data reduction}\label{obs}
\subsection{Instrument, target selection and science observations} \label{InsTagObs}

The work described here is based on data taken with the X-Shooter single target spectrograph mounted on the VLT, Cerro Paranal, Chile \citep{ver11}. The instrument is capable of covering a spectral range which includes UV-Blue (UVB), at 3000-5600 \AA, Visible (VIS), at 5500-10200 \AA, and Near-IR (NIR), at 10200-24800 \AA. A three-arm system is used to observe in all three bands simultaneously at intermediate resolutions (R=3000-17000) depending on the configuration of the instrument. X-Shooter provided the wavelength coverage and spectral resolution required for our science goals. Slit widths of 1.0'', 0.9'', and 0.9'' were used in order to obtain resolutions of R $\sim$ 5100, 8800, and 5100 with pixel scales of approximately 0.1 $\AA$/pix, 0.1 $\AA$/pix, and 0.5 $\AA$/pix  for the UVB, VIS, and NIR arms, respectively. 
Data were collected in November 2009 as part of guaranteed time observation (GTO) program 084.B-0468A. The program was executed in standard nodding mode with an ABBA sequence. Telluric standard stars were taken as part of the GTO program and flux standards were observed through the ESO X-Shooter calibration program and collected from the archive as part of the reduction process. Details of the telluric and flux calibration can be found in the following section. Table \ref{table:obs} lists the cluster names, coordinates, exposure times and signal-to-noise (SNR) values for each arm.  The SNR estimates were obtained by analysing the wavelength ranges of 4550-4750 \AA, 7350-7500 \AA, and 10400-10600 \AA\ for UVB, VIS and NIR arm respectively. \\
The targets in program 084.B-0468A are primarily YMCs in NGC1313 and one in NGC1705, and were selected from a cluster compilation presented by \citet{lar04}. A special effort was made to select YMCs that were isolated and free of contamination from neighbouring objects. Both YMCs studied in this work, along with their host galaxies, are shown in Figure \ref{fig:galaxies}.

\begin{table*}
\caption{X-Shooter Observations}
\label{table:obs}
\centering 
\begin{tabular}{ccccccccc} 
 \hline \hline\\
Cluster & RA & DEC &  & $t_{exp}$ (s)&  & & S/N ( pix$^{-1}$) & \\
 & (J2000) & (J2000) & UVB & VIS & NIR & UVB & VIS & NIR\\
 \hline \\
NGC1313-379 & 49.449038 & -66.50474 & 3120.0 & 3080.0 & 3240.0 & 46.3 & 39.5 & 9.5 \\
NGC1705-1 & 73.55701 & -53.36046 & 600.0 & 580.0 & 600.0 & 144.4 & 108.2 & 39.6 \\
\hline 
\end{tabular}
\end{table*}

\subsection{Data reduction}\label{DR}
Basic data reduction was performed using the public release of the X-Shooter pipeline v2.5.2 and the ESO Recipe Execution Tool (EsoRex) v3.11.1. EsoRex allowed for flexible and tailored use of the standard steps up to the production of the two-dimensional (2D) corrected frames. The basic data reduction included bias (for UVB and VIS) and dark (NIR) corrections, flat-fielding, wavelength calibration and sky subtraction. The UVB and VIS science and flux standard frames were reduced using the nodding pipeline recipe (xsh\_scired\_slit\_nod). The NIR exposures were reduced using the offset pipeline recipe (xsh\_scired\_slit\_offset) due to sky background variations during the individual frames. This approach improved the quality of the products and allowed for a better background subtraction. \par

The spectral extraction was done using the IDL routines developed by \citet{che14} and \citet{gon16}. These routines are based on the optimal extraction algorithm described in \citet{hor86}. The IDL code creates a normalised and smoothed (mean for UVB/VIS and median for NIR) spatial profile,  which is then used alongside an improved version of the bad pixel mask to correct and extract the spectra. After the extraction, the individual orders are combined using a variance-weighted average of each overlapping region.\par

The flux calibration was done using flux standard star observations taken as part of the ESO X-Shooter calibration program. Three different spectrophotometric standards were observed close in time to the science observations, BD+17, GD153, and Feige 110. The observations were made using the 5.0'' wide slit in offset mode. In order to flux calibrate the science exposures, we created response curves for each of the science frames using the pipeline recipe xsh\_respon\_slit\_offset. Offset data reduced in such a mode often provide poor sky corrected products for the NIR frames. For those NIR standard flux observations the pipeline recipe xsh\_respon\_slit\_stare was used instead. This recipe fits the sky on the frame itself, and allows for a better sky correction. All response curves were corrected for exposure time and atmospheric extinction, and were created using the same set of master bias, and master flat field frames as the ones used in their corresponding science exposure. The latter was done in order to correct for the flat-field features varying in time and that remain present in the normalised flat-field images.  After some examination of the response curves we determined that the best and most stable response curve was obtained by using the Feige 110 observations. The criteria for this selection included the rejection of those response curves displaying features or bumps unrelated to the instrument itself. Based on this requirement, Feige 110 was then chosen as the spectrophotometric standard to calibrate all of our science data.  \par

A new response curve was created for each science spectrum. These response curves were used to flux calibrate the science exposures using the following formula:

\begin{equation}
F_\mathrm{cal}(\lambda)=\frac{F_\mathrm{ADU}(\lambda)\ R_\mathrm{cur}(\lambda)\ G\ 10^{(\frac{2}{5} E_\mathrm{c} (\lambda)A_\mathrm{m})}}{E_\mathrm{T}}
\end{equation}

\begin{figure*}
   \resizebox{\hsize}{!}
            {\includegraphics[width=11.2cm]{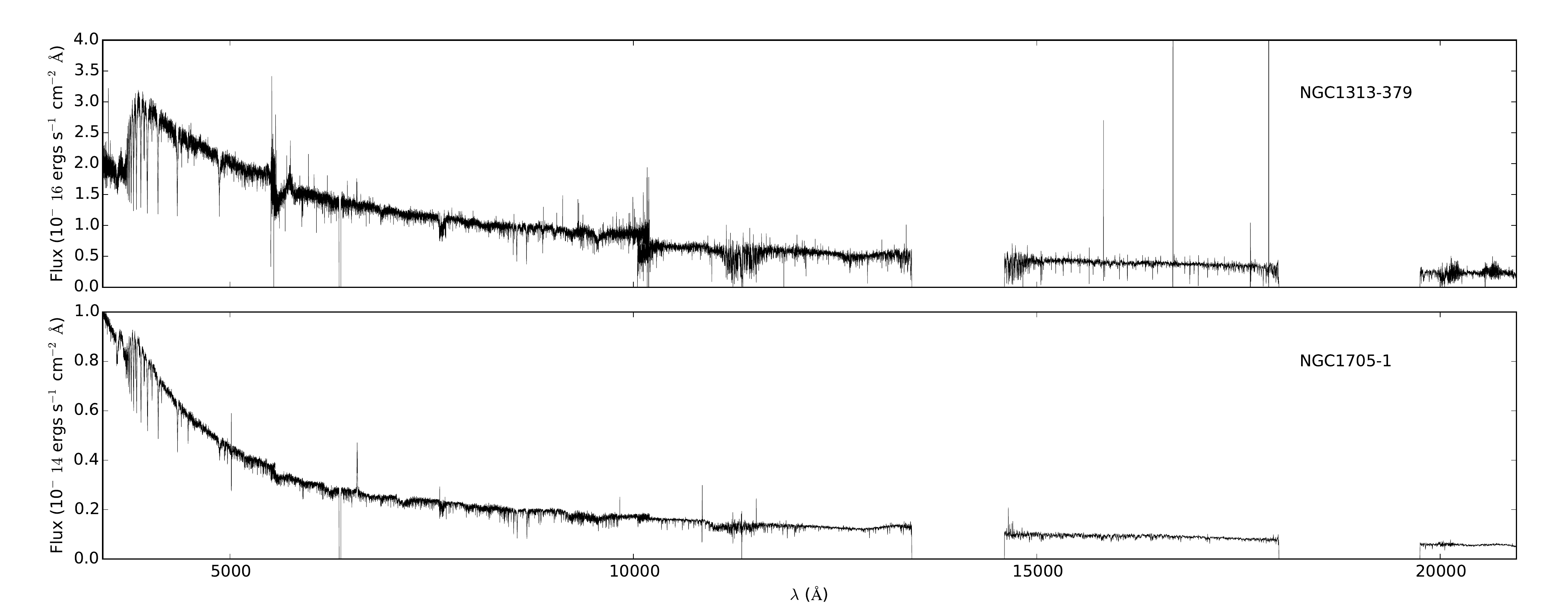}}
      \caption{Example of fully calibrated X-Shooter exposures of NGC1313-379 (top) and NGC1705-1 (bottom). For the benefit of visualisation we partially omitted the noisy edges. We note that a strong diachronic feature is present in the VIS arm below 5700 $\AA$ in the observations for NGC1313-379. See text for more details.   }
         \label{Fig:SpecN1705}
   \end{figure*}

\noindent In the expression above, $F_\mathrm{ADU}$ is the extracted science spectrum, $R_\mathrm{cur}$ is the derived response curve, $G$ is the gain of the instrument (e-/ADU) for the corresponding arm, $E_\mathrm{c}$ is the atmospheric extinction,  $A_\mathrm{m}$ is the airmass, and $E_\mathrm{T}$ is the total exposure time. We note that the slit used when observing the flux standard stars is wider than the one used for the science observations. \citet{che14} pointed out that targets without wide-slit flux correction, requiring a wide-slit science exposure, may lose flux especially in the UVB arm. We visually inspected the calibrated observations and found a good flux agreement between the different arms, we did not observe any obvious flux losses in the UVB exposures (see Figure \ref{Fig:SpecN1705} for an example). We point out that in the observations for NGC1313-379 we see a strong dichroic feature in the VIS arm, below 5700 $\AA$ (Figure \ref{Fig:SpecN1705}, top panel). As noted by \cite{che14}, these features tend to appear at different positions in the extracted 1D data making it difficult to completely remove from the final calibrated spectra. In this case, we have excluded the affected region from our analysis. \par

\begin{figure*}
   \resizebox{\hsize}{!}
            {\includegraphics[width=11.2cm]{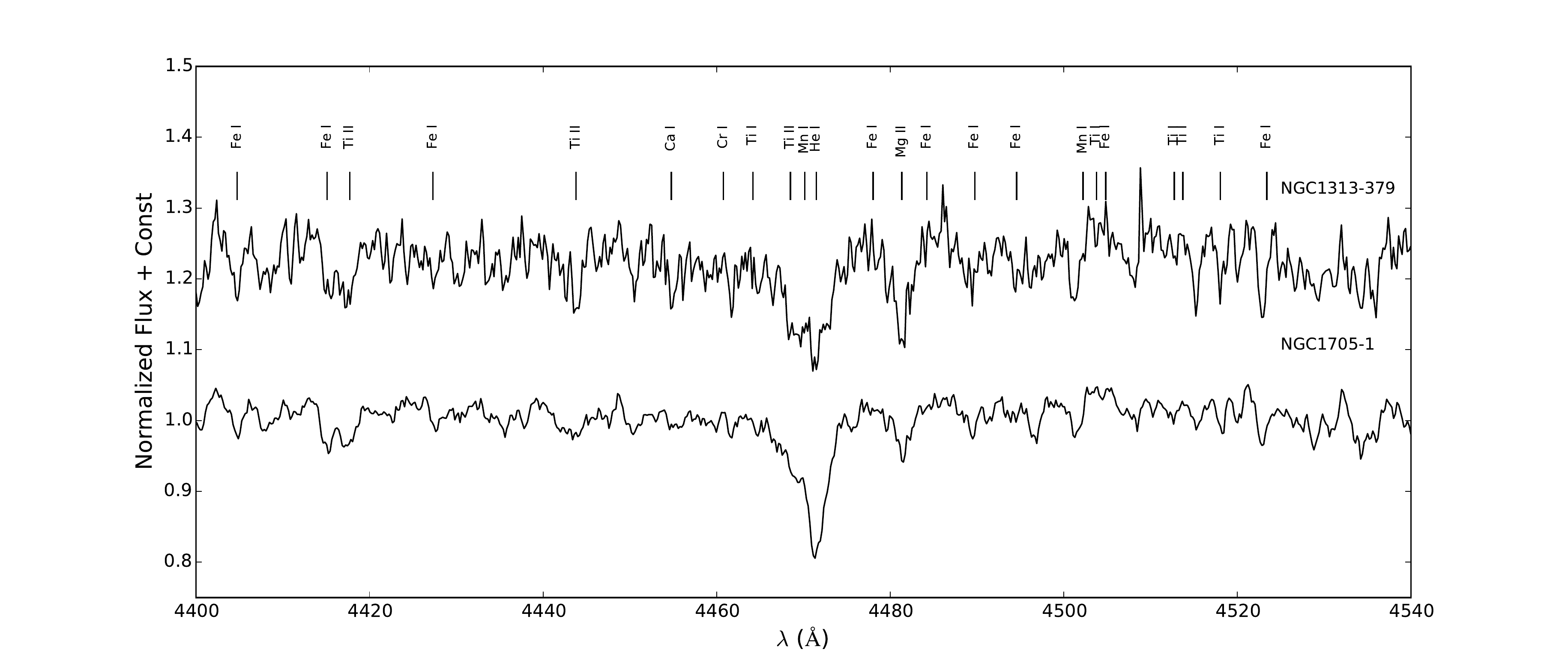}}
      \caption{Sections of the spectra from each of the YMCs analysed in this work. The spectra are normalised to an average continuum level of 1.0, and a constant offset has been added for the benefit of visualisation. }
         \label{Fig:SpecALL}
   \end{figure*}

Ground-based spectroscopic observations are subject to atmospheric contamination. Telluric absorption  features are created by the Earth's atmosphere. These absorption bands, which are mainly generated by water vapor, methane and molecular oxygen, strongly affect the VIS and NIR exposures. We made use of the routines and telluric library created by the X-Shooter Spectral Library (XSL) team to correct for this contamination. For the VIS processing \citet{che14} developed a method based on Principal Component Analysis (PCA) which reconstructs and removes the strongest telluric absorptions. The \citeauthor{che14} method depends on a carefully designed telluric library containing 152 spectra. The NIR telluric correction was done using  routines written by \citet{gon16}. The NIR routines make use of a telluric transmission spectra library known as the Cerro Paranal Advanced Sky Model. The telluric correction for NIR data is done by identifying the best telluric model through PCA and performing a $\chi^{2}$ minimisation to obtain a template of the telluric components included in the science observation. In Figure \ref{Fig:SpecALL} we show a section of the calibrated products for each of the YMCs analysed in this work.  

\section{Abundance Analysis}\label{ana}
We use the L12 approach for analysing integrated-light  observations to obtain detailed abundances, designed and tested on high-dispersion spectra of globular clusters in the Fornax dwarf galaxy. The basic method has been described in detail in \citet{lar12} and \citet{lar14}. Briefly, a series of high-spectral-resolution SSP models are created which include every evolutionary stage present in the star cluster. As a first step we compute atmospheric models using ATLAS9 \citep{kur70} for stars with $T_\mathrm{eff} >$ 3500 K and MARCS \citep{gusta08} for stars with $T_\mathrm{eff} < $3500 K. We opt for using two different sets of models as each is optimised for different temperature ranges. As part of this initial step we specify the stellar abundances for the whole cluster. These atmospheric models are then used to create synthetic spectra with SYNTHE \citep{kur79,kur81} for ATLAS9 models and TURBOSPECTRUM \citep{plez12} for MARCS models. Individual spectra are generated for each of the stars in the cluster and then co-added to produce a synthetic integrated-light spectrum. Before comparing the model spectrum to the science observations, the synthetic data is smoothed to match the resolution of the observations in question. A direct comparison is made between the model spectrum and the science observations, and the process is repeated modifying the abundances accordingly until the best model is determined through the minimisation of $\chi^2$. The current software allows for an overall scaling of all abundances relative to Solar composition \citep{gre98}. This relative scaling parameter is a good approximation to the overall metallicity [m/H] for each of the systems. 

\subsection{ATLAS9 and MARCS models} \label{atmmodels}
ATLAS9 is a local thermodynamic equilibrium (LTE)  one-dimensional (1-D) plane-parallel atmosphere modelling code created and maintained by Robert Kurucz. In order to reduce the computational time, ATLAS9 uses opacity distribution functions (ODFs) when calculating line opacities. New lists of pretabulated  line opacities as a function of temperature and gas pressure for a given wavelength have been created and made available to users for both solar-scaled and $\alpha$-enhanced abundances \citep{cas_kur03}. The calculations presented in this work are based on these new ODFs with solar-scaled abundances. For the synthetic spectral computation of ATLAS9 models we use SYNTHE with atomic and molecular linelists found at the Castelli website \footnote{http://wwwuser.oats.inaf.it/castelli/}. These lists were originally compiled by  \citet{kur90} and later updated by \citet{cas_hub04}. We make use of the Linux versions of ATLAS9 and SYNTHE, and compile the software using Intel Fortran Compiler 12.0.\\
The MARCS models have been developed and focused primarily on constructing late-type model atmospheres \citep{gusta08}. The atmospheric models are LTE 1D plane-parallel or spherical. In contrast to ODFs (as used by ATLAS9), MARCS uses opacity sampling (OS), which treats the absorption at each monochromatic wavelength point in full detail, requiring a large number of wavelength points ($10^5$). For this work we use precomputed atmospheric models obtained from the MARCS website \footnote{http://marcs.astro.uu.se}. Further details of the atomic and molecular spectral line data used in the creation of these models are given by \citet{gusta08}.\par

We note that our analysis is entirely based on LTE modelling and we do not apply any corrections for possible non-LTE (NLTE) effects. As pointed out in \citet{lar14} such NLTE corrections are complicated for any integrated-light studies. Adjustments for NLTE effects depend on the individual star's $\log g$, T$_\mathrm{eff}$ and metallicity. \citet{ber12} estimate corrections (in the J band) for RSGs with metallicities of [Fe/H] > $-$3.5 to be on the order of $<0.1$ dex. For some $\alpha$-elements, such as Mg I, NLTE RSG corrections are higher, varying between $-0.4$ and $-0.1$ dex \citep{ber15}. NLTE line formation might introduce uncertainties that we cannot quantify in this work, however we are aware that this limitation is common in integrated-light studies of star clusters.

\begin{table*}
\caption{Young Massive Cluster Properties}
\label{table:prop}
 \centering 
\begin{tabular}{ccccccc} 
 \hline  \hline \
Cluster & Distance$^a$& log(age) &Metallicity & Mass$^b$ & $M_\mathrm{TO} $ & $M_\mathrm{Giants}$\\
 & [Mpc] &  & (Z) & [$M_{\odot}$] & [$M_{\odot}$] & [$M_{\odot}$]\\
  \hline\\ 
NGC1313-379 & 4.1 & $7.74^c$  &0.004& $2.8 \times 10^5$& 6.38 & 6.50\\
NGC1705-1 & 5.1 & $7.10^d$& 0.008&$9.2 \times 10^5$& 15.34& 15.58\\
 \hline 
 \end{tabular}
\\ \textsuperscript{$a$} \textit{NASA/IPAC Extragalactic Database (NED)}, \textsuperscript{$b$} \citet{lar11}, \textsuperscript{$c$} \citet{lar99}, \\ \textsuperscript{$d$} \citet{vaz04}
\end{table*}

\subsection{Stellar parameters} \label{stel_par}
In order to create or select an atmospheric model, we first generate a HRD, covering every evolutionary stage present, for each of the YMCs studied in this work. Both clusters discussed in this paper have published CMDs available from HST observations. \par
\begin{figure*}
\centering
\includegraphics[width=9.0cm]{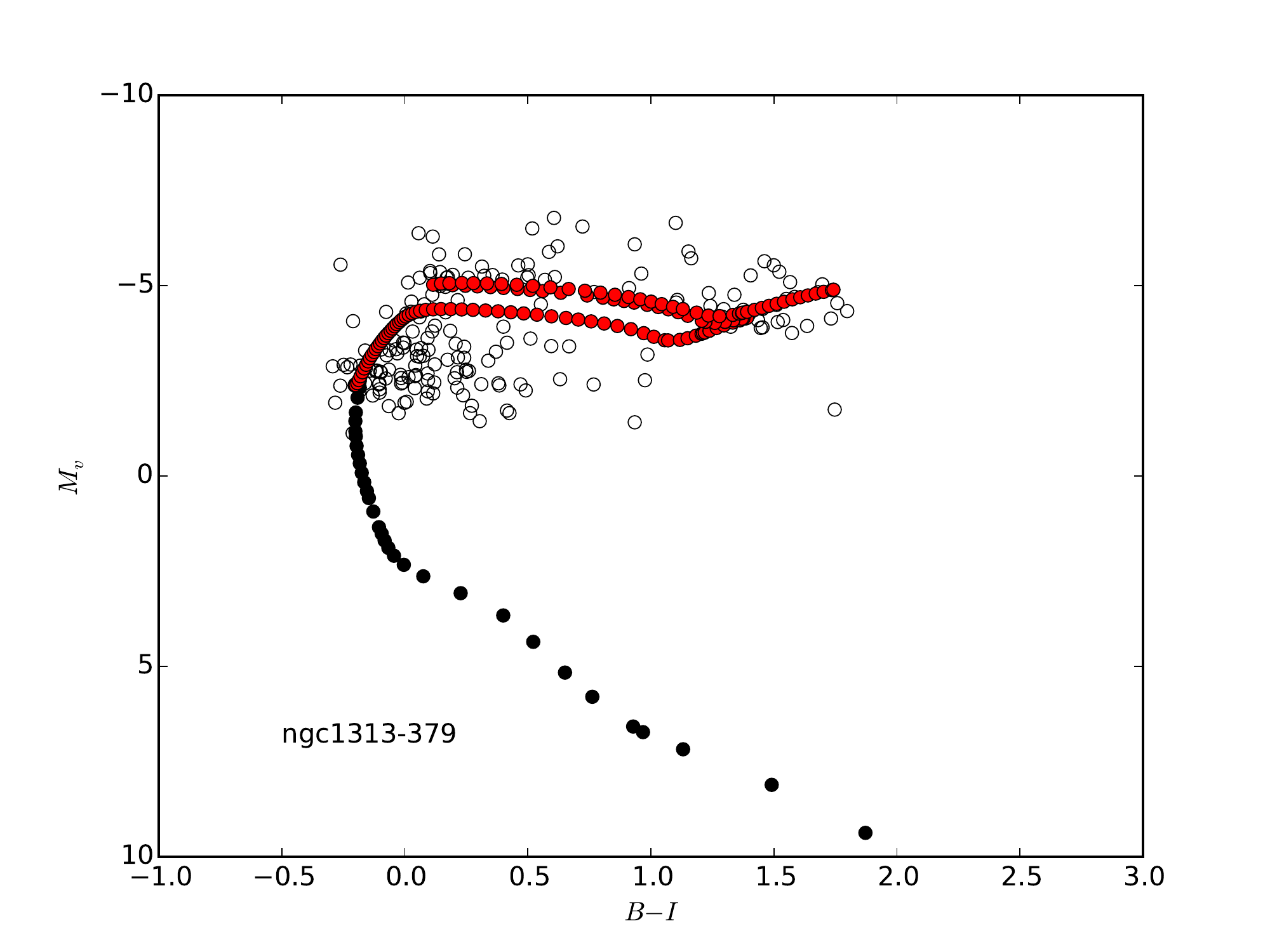} 
\includegraphics[width=9.0cm]{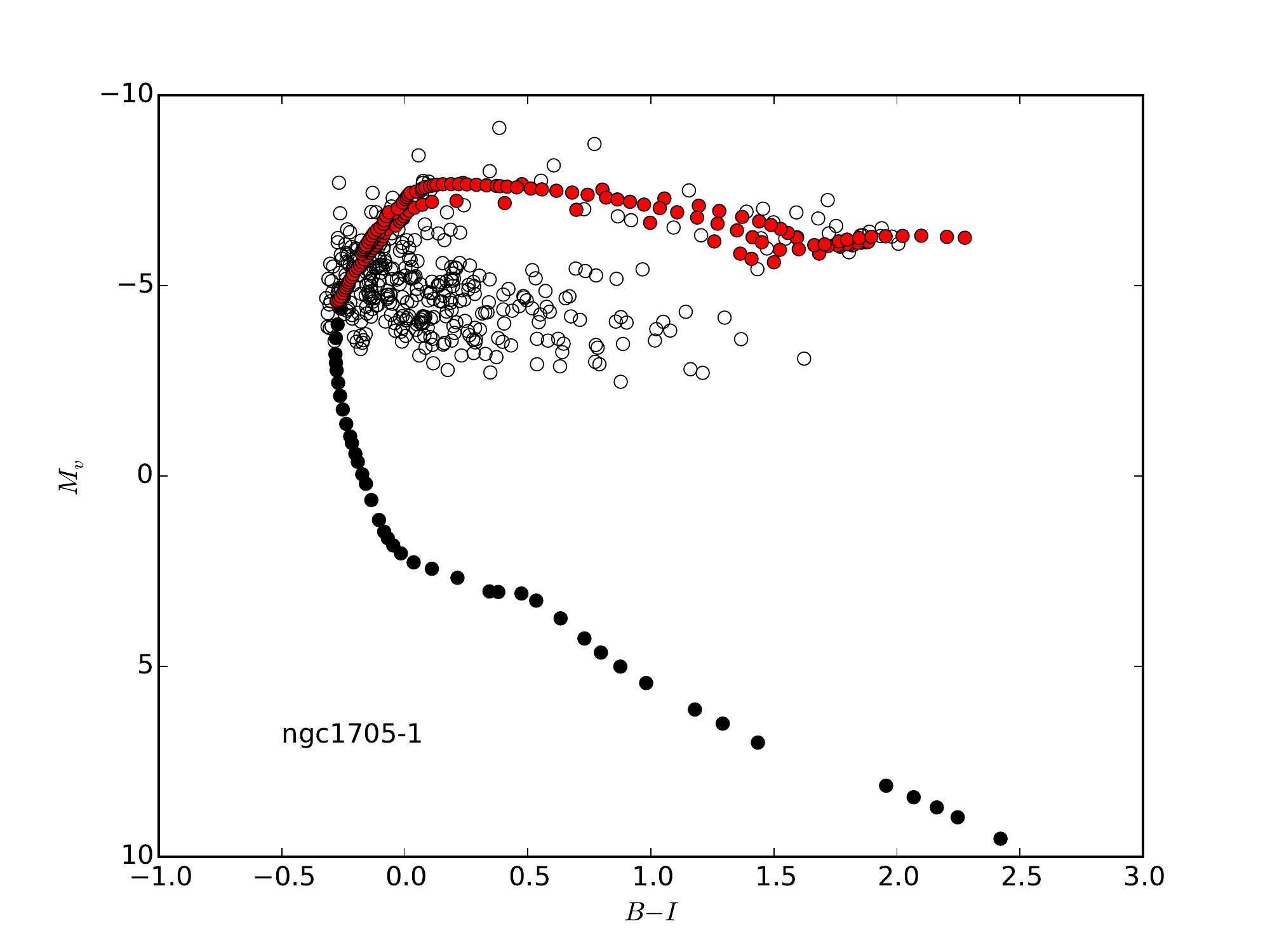}
\caption{CMDs used in order to estimate the stellar parameters for model atmosphere and synthetic integrated-light spectra. Filled circles, represent the colours extracted from theoretical isochrones for metallicities Z=0.004 and Z=0.008, and ages of 56 Myr and 12 Myr for NGC1313-379 and NGC1705-1, respectively. Empty circles show the empirical CMDs from \citet{lar11}. In the Isochrone-Only method we use both, red and black filled circles. For the CMD+Isochrone method we instead use the empty circles and the black filled circles.}
\label{fig:cmds}
\end{figure*}

However, the observations only cover the most luminous stars in the clusters, mainly those brighter than the main sequence turn-off. Due to this incompleteness we use the CMDs to estimate the stellar parameters of the brightest stars, constraining the location of the red/blue supergiants on the HRD and obtaining a scaling of the total number of stars within the cluster (see Figure ~\ref{fig:cmds}). \par
We use the calibrated CMDs from \citet{lar11} and convert the ACS instrumental magnitudes to standard Johnson-Cousins \textit{V }and \textit{I} magnitudes using the synthetic photometry package PySynphot. We derive the $T_\mathrm{eff}$ and bolometric corrections from the \textit{V-I} colours using the Kurucz colour-$T_\mathrm{eff}$ transformations. Kurucz colour tables have lower ($\sim -$0.32) and upper ($\sim +$2.80) limits in $V-I$, therefore any stars in the empirical data with $V-I$ values outside these limits are excluded from the analysis. For a discussion on systematic uncertainties arising from the $T_\mathrm{eff}$ obtained through the \textit{V-I} colours, we refer to Section ~\ref{Teff}.\par

The surface gravities, $\log g$, are estimated using the simple relation described in equation (1) of L12. For individual stars in the empirical data two different masses are assumed. Stars at the main sequence (MS) turn-off (TO) point or below are assigned the MS TO mass ($M_\mathrm{TO}$) for a population of that particular age. For stars above the TO point we use the average mass of the population of giants ($M_\mathrm{Giants}$). Table \ref{table:prop} lists the different assigned masses for the individual clusters.\par
For stars below the detection limit we use theoretical models to obtain the physical parameters. We use PARSEC theoretical isochrones from \citet{bre12} and assume an IMF following a power law, $dN/dM \propto M^{-\alpha}$. We adopt a \citet{sal55} IMF with an exponent of $\alpha=2.35$ and a lower mass limit of 0.4$M_\mathrm{\odot}$. The total number of stars present in the clusters is estimated using the supergiants from the empirical CMD as a scaling factor for the stellar population. We define as a supergiant any objects with values of $\log g \leq$ 1.0. First, a population of 100,000 stars is generated using the theoretical isochrone with a Salpeter exponent. The supergiant ratio of empirical to isochrone stars is then used as a scaling element, $S_\mathrm{cmd/iso}$. The final number of stars ($dN$) for a specific stellar type is defined using the following relation

\begin{equation}
dN= 10^5 \: S_\mathrm{cmd/iso} \: dM \: M^{\alpha}
\end{equation}

\noindent In addition to the physical parameters mentioned above, we also account for the microturbulent velocity, $\nu_{t}$, defined as the non-thermal component of the gas velocity in the region of spectral line formation at small scales \citep{canti09}. As described by L12  we fit for the overall scaling abundance and the microturbulence parameters simultaneously for one of our young clusters, NGC1313-379. Currently, the code fits for a single microturbulent velocity which is then applied to all of the stars in the cluster. This test is done on 200 \AA \ bins covering UVB and VIS/Optical wavelengths, 4000-5200 \AA \ and 6100-9000 \AA, \ respectively. We find a poorly constrained mean microturbulence of $\langle \nu_\mathrm{t} \rangle = 3$ km s$^{-1}$ with large bin-to-bin dispersion of $\sigma(\nu_\mathrm{t})=  1.25 $\ km s$^{-1}$. This value is comparable to an average microturbulence $\langle\nu_\mathrm{t}\rangle = 2.8 $\ km s$^{-1}$ observed in individual RSGs in the SMC \citep{dav15}, and $\langle\nu_\mathrm{t}\rangle = 3.3 $\ km s$^{-1}$ measured in RSGs in the LMC \citep{gaz14}. However, after performing several tests adjusting the microturbulence values between $ \nu_\mathrm{t} = 2$ km s$^{-1}$ and $\nu_\mathrm{t} = 3$ km s$^{-1}$ for stars with T$_\mathrm{eff}$ < 6000 K, we notice that the scatter in the overall metallicity measurements is reduced when using a microturbulence value of $\nu_\mathrm{t}  = 2$ km s$^{-1}$ for both NGC1313-379 and NGC1705-1. This value is comparable to what \citet{lar15} measured for three Super Star Clusters in NGC4038. Additionally, previous studies have shown that microturbulence is correlated to T$_\mathrm{eff}$ \citep{lyu04} and  anticorrelated to $\log g$. The multiple microturbulence values seen in different stellar types mean that using a single value for the whole stellar population is an oversimplification. Instead, we assign microturbulence values as follows: $\nu_\mathrm{t} = 2.0$ km s$^{-1}$ for stars with T$_\mathrm{eff} <$ 6000 K, $\nu_\mathrm{t} = 4.0$ km s$^{-1}$ for stars with 6000 $<$ T$_\mathrm{eff} < $ 22000 K  \citep{lyu04} and $\nu_\mathrm{t} = 8.0$ km s$^{-1}$ for stars with T$_\mathrm{eff} >$ 22000 K \citep{lyu04}. We note that microturbulence for fainter MS stars is observed to be lower than the values we assign as part of this study. One example is the Sun, where studies measure microturbulence values of $\nu_\mathrm{t} \sim 1.0$ km s$^{-1}$ \citep{pav12}. However, we note that the contribution of these types of stars to the total integrated light is relatively small.

\subsection{Matching the resolution of the observations} \label{resol}
\begin{figure}
   \centering
   \includegraphics[scale=0.4]{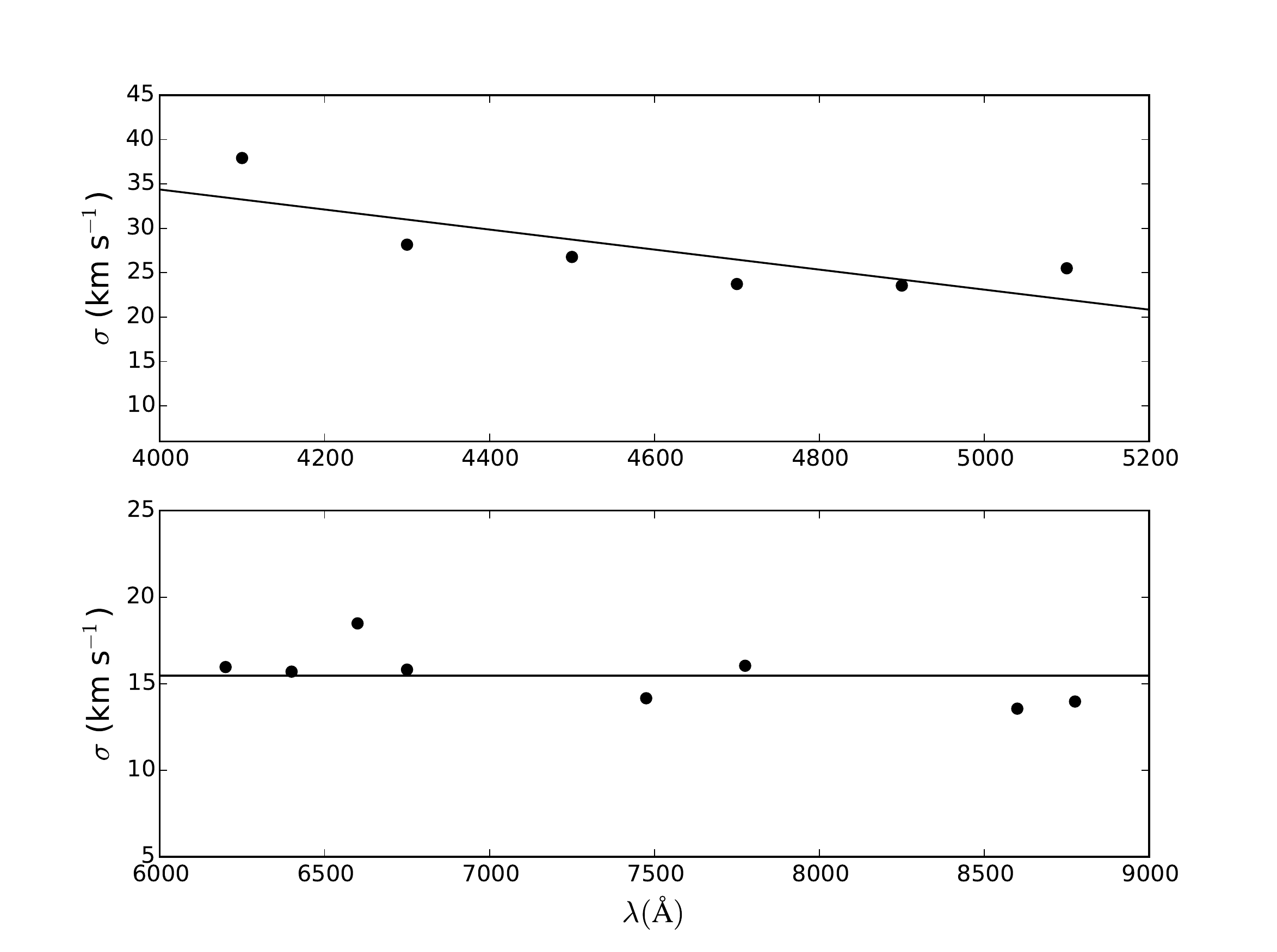}
      \caption{Measured $\sigma_{sm}$ in km s$^{-1}$ as a function of wavelength for NGC1313-379. Top panel shows the $\sigma_{sm}$ obtained in the UVB arm along with a first-order polynomial fit. Bottom panel presents the $\sigma_{sm}$ calculated in the VIS arm with the mean $\sigma_{sm}$ indicated by a black line. 
              }
         \label{fig:sm}
   \end{figure}

The model spectra are created at high resolution, $R=500,000$, and then degraded to match the X-Shooter observations. The current code offers the option of estimating the best-fitting Gaussian dispersion value ($\sigma_{sm}$) to be used in the smoothing of the model. We fit each spectrum in 200 $\AA$ bins, allowing the $\sigma_{sm}$ and [m/H] to vary. This provides us with the best $\sigma_{sm}$ and overall metallicities to be used in the selection of the isochrone. In Figure ~\ref{fig:sm} we show our best $\sigma_{sm}$ for  NGC1313-379 as a function of wavelength for both the UVB and VIS arm. This $\sigma_{sm}$ value accounts mainly for the finite instrumental resolution and internal velocity dispersions in the cluster, and to a lesser degree for the stellar rotation and macroturbulence. The latter component is a spectral line broadening caused by convection in the outer layers of individual stars and does not follow a Gaussian profile. We note that in certain cases, macroturbulence has been measured to be significant with values around $\sim$ 10 km s$^{-1}$ \citep{gra87}, comparable to the velocity dispersions in the clusters themselves.\par

Since X-Shooter collects data through a multi-arm system, we estimate different $\sigma_{sm}$ values for each of the arms. \citet{che14} found that the X-Shooter resolution varies with wavelengths in the UVB arm, and remains constant in the VIS arm. After fitting for the best $\sigma_{sm}$ we also see that the resolution in the UVB arm has a stronger dependance on wavelength than the VIS arm (See Figure ~\ref{fig:sm}). In their study, \citet{che14} used the 0.5" slit for the UVB and 0.7" for the VIS arm. Given that our X-Shooter observations use a different instrument configuration, a direct resolution comparison with  \citet{che14} is not possible. Assuming that the resolution in the VIS arm does not vary with wavelength, and that the resolving power represents a Gaussian full width at half maximum (FWHM), the instrumental resolution for the configuration used  (R=8,800\footnote{https://www.eso.org/sci/facilities/paranal/instruments/xshooter/inst.html}) corresponds to $\sigma_{inst}$ = 14.47 km s$^{-1}$.  The line-of-sight velocity dispersions for the individual clusters are estimated using the best-fitting Gaussian dispersion found for the VIS arm only. 

\subsection{Clean lines} \label{lines}
One of the main challenges in the analysis of integrated-light spectroscopy for detailed abundances is the degree of blending of spectral features. The L12 technique has previously been applied to high-resolution observations of (mainly metal-poor) GCs. The resolution of our data imposes additional challenges as intermediate-resolution observations are more strongly affected by blending. In general, younger star clusters have higher metallicities, which in turn means higher degree of blending. \par
To be able to utilise the L12 method on the X-Shooter data sets, we create optimised wavelength windows tailored for each element. These windows are defined and selected in an attempt to ameliorate the effects of strong blending in regions where element lines overlap with lines of a different species. \par
\begin{figure}
   \centering
   \includegraphics[width=\hsize]{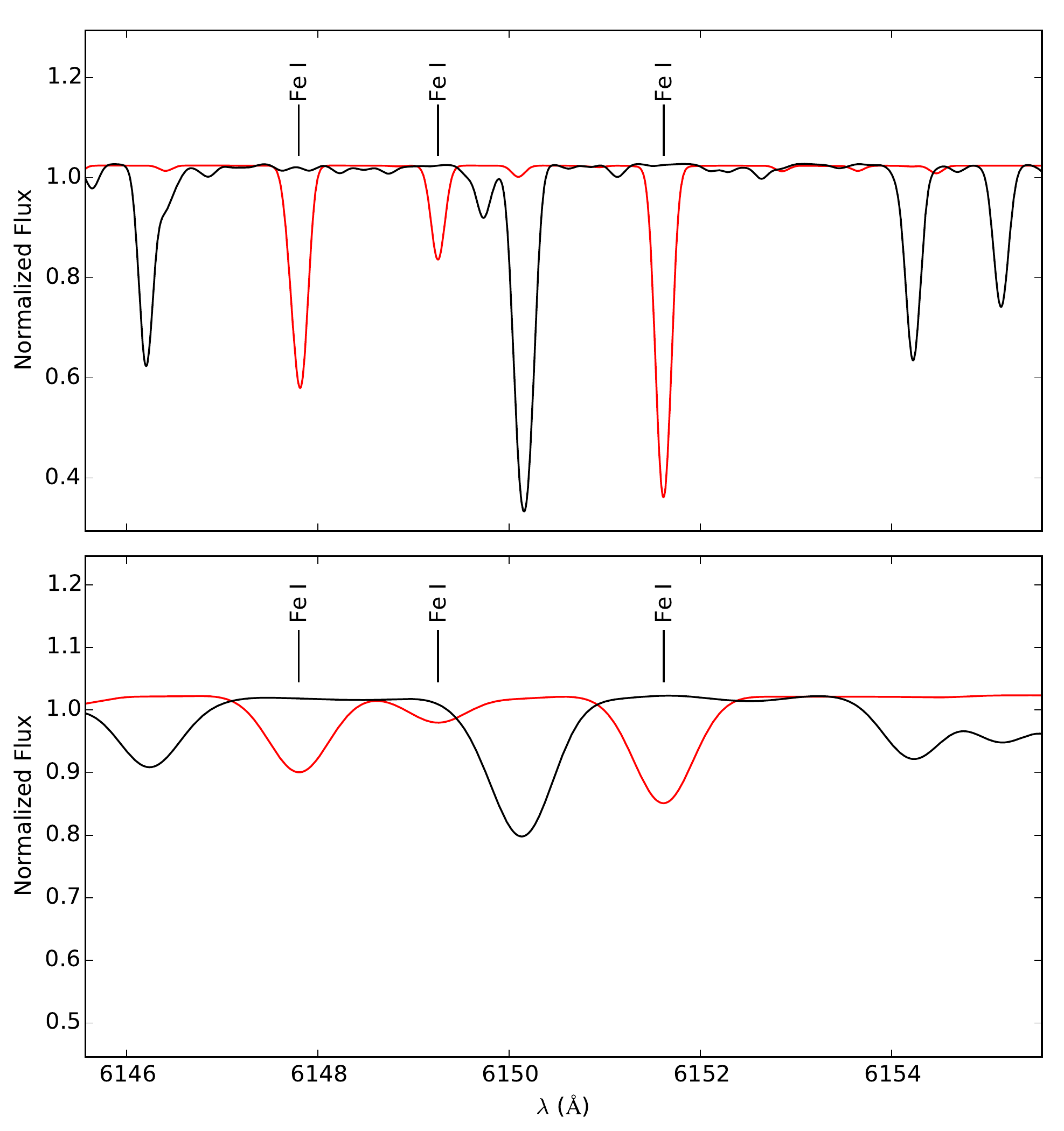}
      \caption{Arcturus synthetic spectra used in the selection of wavelength bins to exclude strong, blended lines. In red the synthetic Arcturus spectrum including only Fe lines. In black the model Arcturus spectra including all line species, except Fe. Top: High-resolution model, R=47,000. Bottom: Low-resolution model, similar to the resolution offered by X-Shooter, R = 8,800, shown just for comparison. The line selection was done using the high-resolution models.
              }
         \label{ArcSpec}
   \end{figure}

 
We generate a stellar model with physical parameters similar to those determined for Arcturus by \citet{ram11}: $T_\mathrm{eff} = 4286\,$K, log $g$ = 1.66, and $R=25.4 \,R_{\odot}$. The model is used to produce two sets of high-resolution (R=47,000) synthetic spectra. The first synthetic spectrum excludes any lines of the element in question, for example, Fe. The second spectrum includes only lines of the element under study. A Python script compares the two spectra and highlights, in this example, Fe lines that are not blended with other elements. For pre-selection/exclusion we enforce the following criteria:
\begin{itemize}
\item The depth of the element line must have a maximum flux of 0.85 (from a normalised spectrum);
\item Any lines of the same element found within $\pm$ 0.25$\AA$ from the line in question are excluded; and
\item Any lines where the flux in the black spectrum (from Figure ~\ref{ArcSpec}, top subplot) is lower than 0.90 (from a normalised spectrum) within  $\pm$ 0.25$\AA$ from the line in question are excluded.
\end{itemize}

An example of this preselection is shown in Figure ~\ref{ArcSpec}, where the spectrum in red belongs to the synthetic spectrum with only Fe lines, compared to the spectrum in black, which includes every element line except Fe. Marked with vertical lines are those species considered clean and unaffected by blending. Once the code preselects clean lines based on the criteria described above, we then visually inspect the highlighted lines and define wavelength windows that include these clean elements.  This procedure is repeated for every element included in this work. We point out that for Fe and Ti these windows are broader and cover more than one element line (see Tables~\ref{table:chem379} and ~\ref{table:chem1}). These specific windows are chosen due to the large number of Fe and Ti lines found throughout the spectral range, but are selected using our clean-line technique in an effort to include as many lines as possible. As expected for such broad windows, these include several blended lines.  

Stars cooler than Arcturus (T$_\mathrm{eff}<4200$K) are strongly affected by TiO bands. We remark that the window selection described above is based on model spectra of Arcturus and is focused on atomic lines. In general this selection does not exclude regions where TiO blanketing is present, however these molecular lines are included in our standard synthesis line lists. Furthermore, the strength of these bands is highly sensitive to metallicity; higher metallicity leads to stronger TiO absorption \citep{dav13}. To test how sensitive our abundance measurements are to these TiO bands, we estimate the abundances with and without TiO line lists. We find differences in the measured abundances of < 0.03 dex further confirming that the impact of TiO blanketing is small at this metallicity.

\subsection{Individual clusters} \label{ind_cl}
\subsection*{NGC1313-379}
NGC1313-379 is a YMC with an approximate age of 56 Myrs and a mass of $2.8 \times 10^5$$M_{\odot}$ \citep{lar11}.  From spectroscopy of H II regions, \citet{wal97} found an oxygen abundance of 12 +log O/H = 8.4 for NGC1313, which is similar to the oxygen abundance for young stellar populations and H II regions in the LMC \citep{rus92}. The H II region abundances in NGC1313 show no significant radial gradient across the disk. The analysis for this YMC is first done combining the HST CMDs and theoretical isochrones as described above. For those stars below the detection limit we initially use an isochrone of log(age)=7.75 and metallicity Z=0.007. With this combination of CMD+Isochrone we measure an overall metallicity of [m/H] $\sim$ $-$0.6 dex. For our final analysis we combine the HST CMD with an isochrone of log(age)=7.75 and metallicity Z=0.004. \par
In order to test the sensitivity of our results to the CMD+Isochrone method, we repeat the analysis using only the isochrone to set up the stellar parameters (from now on we refer to it as Isochrone-Only method). When comparing the abundances obtained with the CMD+Isochrone method to the ones calculated with the Isochrone-Only method, using an isochrone log(age)=7.75 and Z=0.004 for both procedures, we find that the differences are $\leq$0.1 dex. More details on the differences on individual abundances obtained using the two different methods and the uncertainties that might be introduced by each of them are discussed and displayed in Section ~\ref{sens_iso} and Table ~\ref{table:iso_sensitivity}. \par
The best-fitting Gaussian dispersion for the smoothing of the VIS synthetic spectra is $\sigma_{sm}$ = 15.5 $\pm$ 1.5 km s$^{-1}$. Subtracting the instrumental broadening in quadrature, we find a line-of-sight velocity dispersion for the cluster of $\sigma_{1D}$ = 5.4 $\pm$ 4.2 km s$^{-1}$. 

\subsection*{NGC1705-1}
A mass of $9.2 \times 10^5$ $M_{\odot}$ \citep{lar11} and an age of 12 Myr \citep{vaz04} makes this the younger and more massive YMC in this study. H II regions in NGC1705 have been found to have a metallicity similar to the young stellar component in the SMC \citep{lee04}. As a first test we take the CMD from \citet{lar11} as published, and combine it with an isochrone of log(age)=7.1 and metallicity Z=0.008. The age and metallicity of the isochrone used in the initial trial are chosen based on the CMD simulation parameters assumed in \citet{lar11}. We measure the overall metallicity of the cluster using 200 \AA \ bins (as described in Section ~\ref{resol}) and find a rather strong trend with wavelength. This same trend is also observed in our measurements of Ti. \citet{lar11} points out that the simulated CMD for this specific YMC was a poor match to the HST observations. An additional mention is made regarding the observed colours of the red supergiant stars, which tend to be redder in the observations of NGC1705-1 than predicted by the models. \citet{lar11} used Padua isochrones \citep{ber09}, but the overall metallicity trend observed in this work and the mismatch between model and HST observations in \citet{lar11} can be understood as further confirmation of previously known issues found in canonical stellar isochrones (see \citealt{lar11} and \citealt{dav13} for a detailed discussion). Due to the metallicity trends observed when using CMD+Isochrone, we opt for proceeding with an analysis where all the stellar parameters are extracted from the theoretical isochrone alone.\par
\begin{figure}
   \centering
   \includegraphics[width=\hsize]{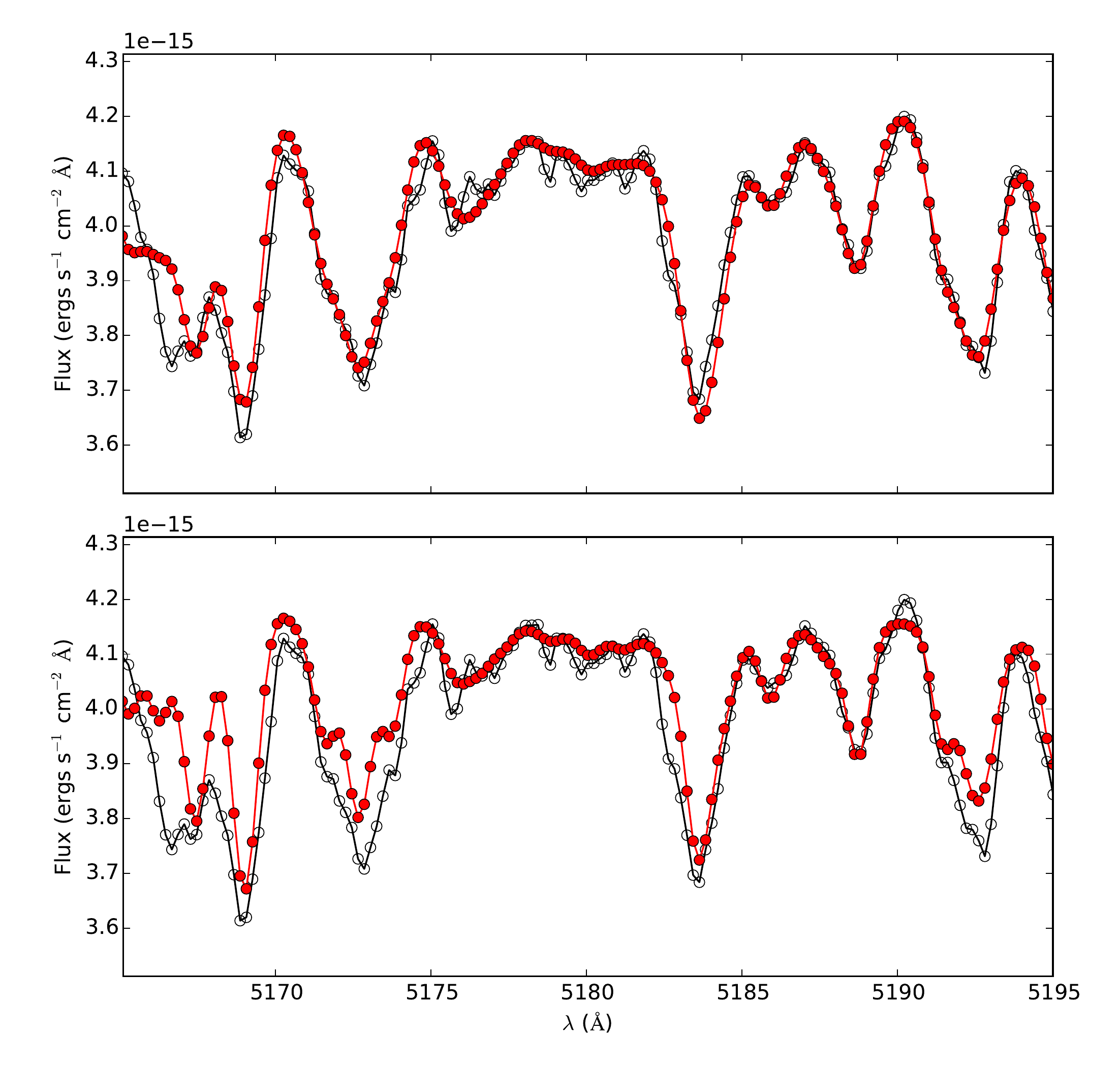}
      \caption{Top: Best model spectra for YMCs NGC1705-1 using an isochrone of log(t)=7.75 and Z=0.008. Bottom: Best model spectra for the same YMCs as above using an isochrone of log(t)=7.75 and Z=0.004. The black curve represents the X-Shooter science observations and red points show the best model spectra. 
              }
         \label{Fig:1705_comp}
   \end{figure}
\begin{figure*}
   \centering
   \includegraphics[width=\hsize]{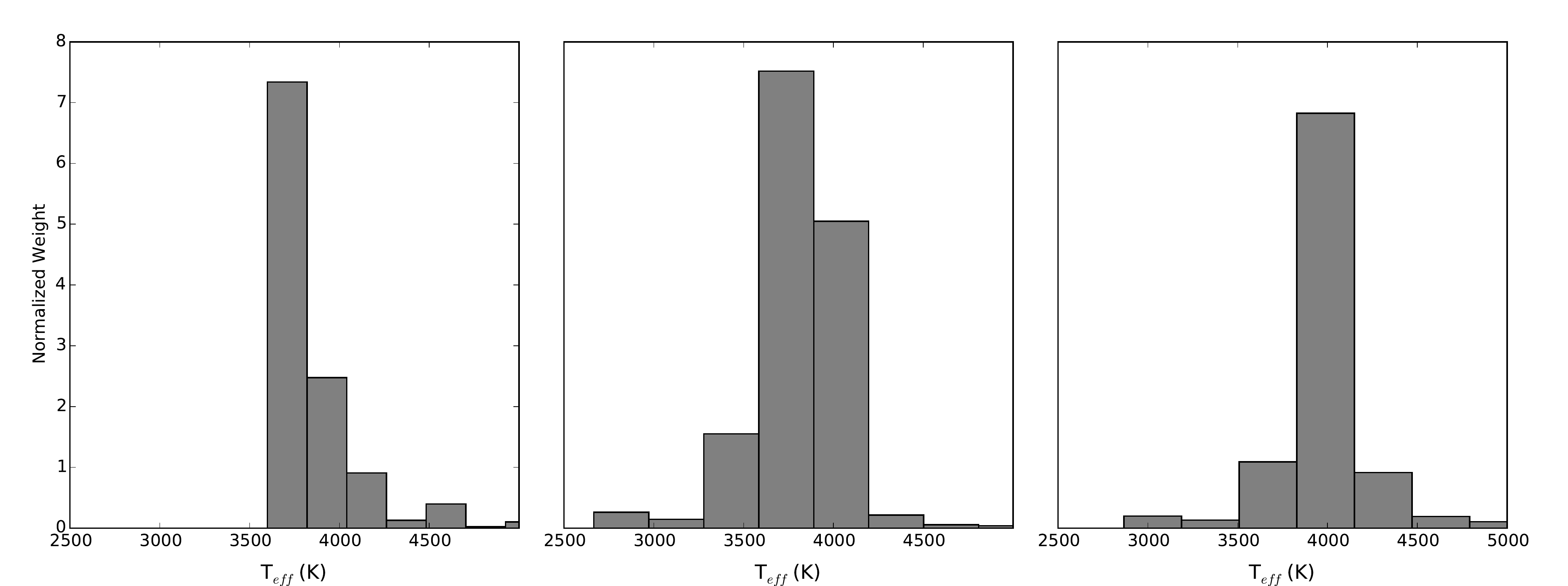}
      \caption{Weight distribution as a function of T$_\mathrm{eff}$ for NGC1705-1. Left panel: stellar parameters extracted from the empirical CMD, with a peak around $\sim$3700 K. Middle panel: stellar parameters obtained from the theoretical isochrone for log(age)=7.1 and Z=0.008, the distribution peaks at $\sim$3700 K. Right panel: stellar parameters extracted from the theoretical isochrone for log(age)=7.1 and Z=0.004, the peak is located at $\sim$4000 K.
              }
         \label{Fig:histo_1705}
   \end{figure*}

Our first estimate gives an overall metallicity of [m/H]  $\sim -$0.78 dex. This metallicity is substantially lower than the isochrone used (Z=0.008 or [m/H] = $-$0.28). To preserve self-consistency, we change the isochrone metallicity to Z = 0.004. This second estimate is consistent with the initially inferred metallicity; however, after visually inspecting the individual fits we realise that the best model spectra generated using the lower metallicity isochrone (Z=0.004) do not match the observations as well as the higher-metallicity isochrone (Z=0.008); see Figure ~\ref{Fig:1705_comp} for a comparison of the best model spectra obtained with the different metallicity isochrones. In addition, we obtain a lower reduced-$\chi^2$ with the higher-metallicity isochrone. These differences in the results appear to be arising from the effective temperature (T$_\mathrm{eff}$) distribution in the different theoretical isochrones. In Figure ~\ref{Fig:histo_1705} we show the distribution of weights as a function of T$_\mathrm{eff}$ for the empirical data (CMD - left plot), isochrone Z= 0.008 (middle plot), and isochrone Z=0.004 (right plot). We define the weight of the different types of stars as 

\begin{equation}
w = n_\mathrm{star} \: R_\mathrm{star}^{2}
\end{equation}

\noindent where n$_\mathrm{star}$ is the total number of stars for the corresponding temperature, and R$_\mathrm{star}$ is the radius of the star. From this comparison we can see that the weight in the empirical CMD peaks at $\sim$ 3700 K, similar to what is observed in the high-metallicity isochrone (Z=0.008). On the other hand, for the lower-metallicity isochrone (Z=0.004), the weight distribution reaches a maximum at temperatures around $\sim$ 4000 K. This comparison shows that the temperature distribution in the higher metallicity isochrone best represents the distribution of temperatures observed in the CMD. This difference in temperature distributions could also be the cause of the discrepancies observed in the best model spectra shown in Figure ~\ref{Fig:1705_comp}. The final analysis is done using the Isochrone-Only method, extracting the stellar parameters from the isochrone with log(age)=7.1 and Z=0.008.\par
We find a best-fitting smoothing of $\sigma_\mathrm{sm}$ = 16.7 $\pm$ 1.7 km s$^{-1}$, and a line-of-sight velocity dispersion of about $\sigma_\mathrm{1D}$ = 8.3 $\pm$ 3.4 km s$^{-1}$, slightly higher than what is estimated for NGC1313-379 but expected as NGC1705-1 is more massive. For this YMC, \citet{hof} found a line-of-sight stellar velocity dispersion of $\sigma_\mathrm{1D}$ = 11.4 $\pm$ 1.5 km s$^{-1}$, consistent within the errors of our measured velocity dispersion. 

\section{Results}\label{results}
We measure a number of individual elements from the YMC spectra starting with those having the highest number of lines. As a first step, we fit for the smoothing parameter and overall metallicity factor, [m/H], scanning the whole wavelength range 200 $\AA$ at a time. The continua of the model and observed spectra are matched using a cubic spline with three knots. The overall metallicity values estimated for the 4000-4200 $\AA$ and 4200-4400 $\AA$ bins are consistently lower than the rest of the bins with differences of  $\leq$0.4 dex. This behaviour is true for both YMCs. We suspect that this is caused by heavy blending (due to higher-metallicities than those observed in GCs) at these young ages. Given the broad wavelength coverage that we have available with X-Shooter, the exclusion of these problematic bins does not impact our analysis. We then proceed keeping the overall scaling and the smoothing parameters fixed. Fe is the first element we measure since this element has the largest number of lines populating the spectra, followed by Ti and Ca. Using the tailored wavelength windows described in Section \ref{lines}, we obtain individual abundance measurements for NGC1313-379 and NGC1705-1, which we list in Tables \ref{table:chem379} and \ref{table:chem1}, respectively. Individual elements, wavelength bins, best-fit abundances and their corresponding 1-$\sigma$ uncertainties calculated from the $\chi^2$ fit are shown in these tables. We use a cubic spline with three knots to match the continua of the model and observed spectra for those windows $\geq$ 100 $\AA$. For bins narrower than 100 $\AA$ we use a first-order polynomial instead. \par

   \begin{figure}
   \centering
   \includegraphics[scale=0.50]{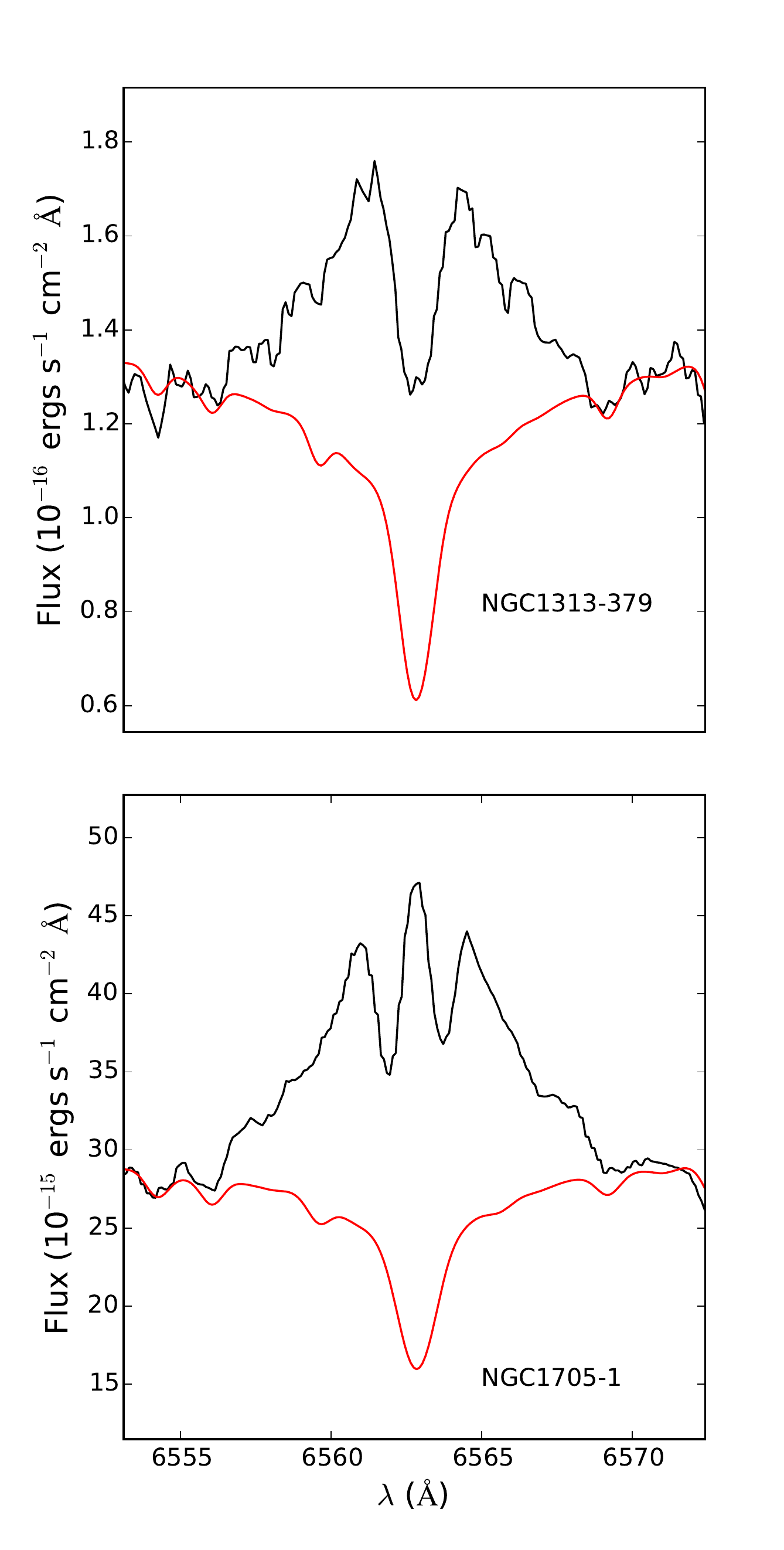}
      \caption{H$\alpha$ emission lines in NGC1313-379 (top) and NGC1705-1 (bottom). In black we show the X-Shooter observations, and in red we display the model spectra. 
              }
         \label{Fig:balmer}
   \end{figure}

   \begin{figure}
   \centering
            {\includegraphics[scale=0.55]{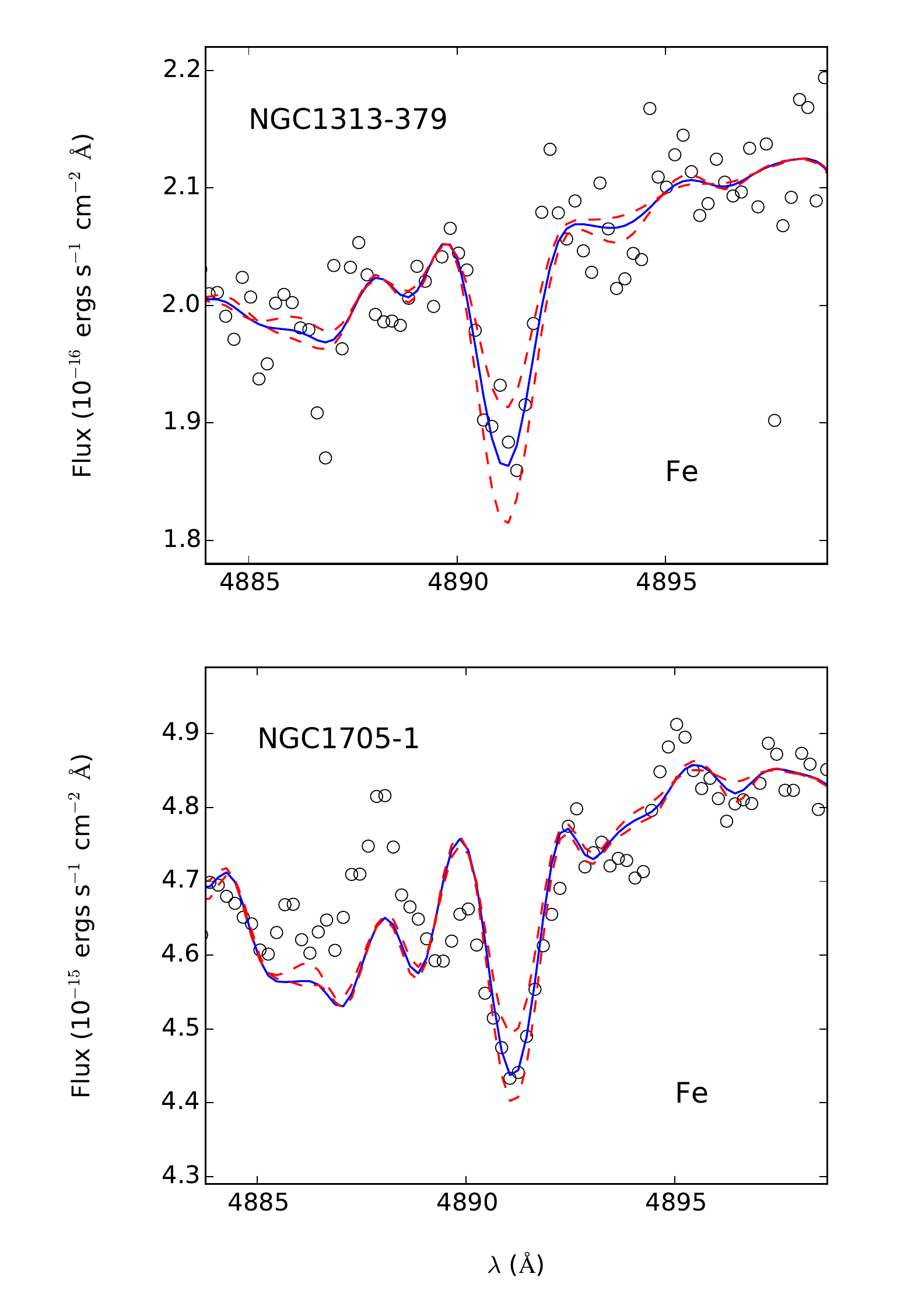}}
      \caption{Example synthesis fits of an Fe I line in our X-Shooter spectra of NGC1313-379 (top) and NGC1705-1. Empty black circles are the X-Shooter observations. Blue curve shows the best abundance. Red dashed curves show the effect of varying the element in question by $\pm$0.5 dex. }
         \label{Fig:Fe}
   \end{figure}

   \begin{figure*}
   \resizebox{\hsize}{!}
            {\includegraphics[width=11.2cm]{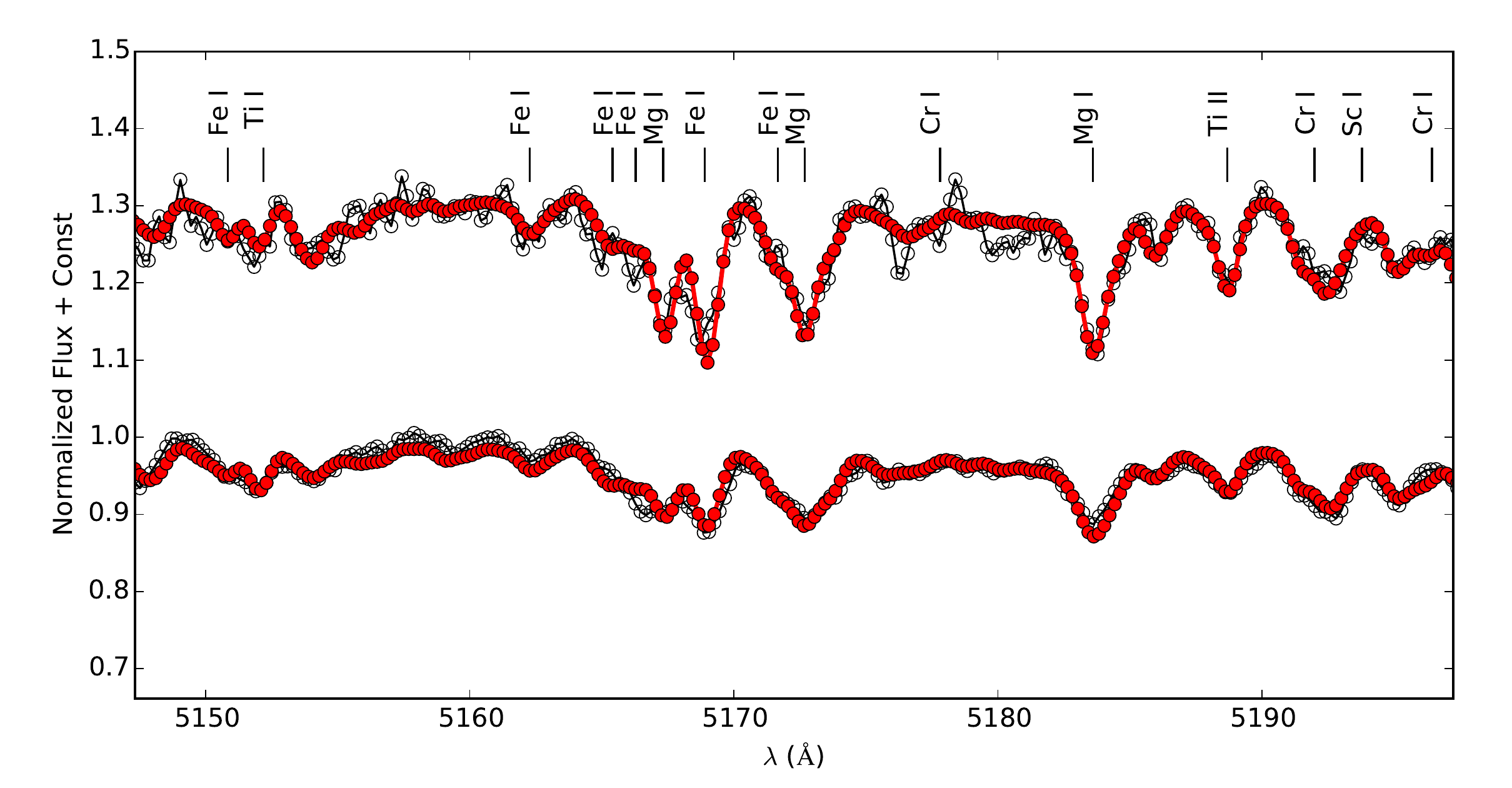}}
      \caption{Example synthesis fits for NGC1313-379 (top) and NGC1705-1 (bottom). Empty black circles correspond to the X-Shooter observations, while the filled red circles are the best-fitting model spectra.}
         \label{Fig:Spec1317}
   \end{figure*}
   \begin{figure}
   \resizebox{\hsize}{!}
            {\includegraphics[width=11.2cm]{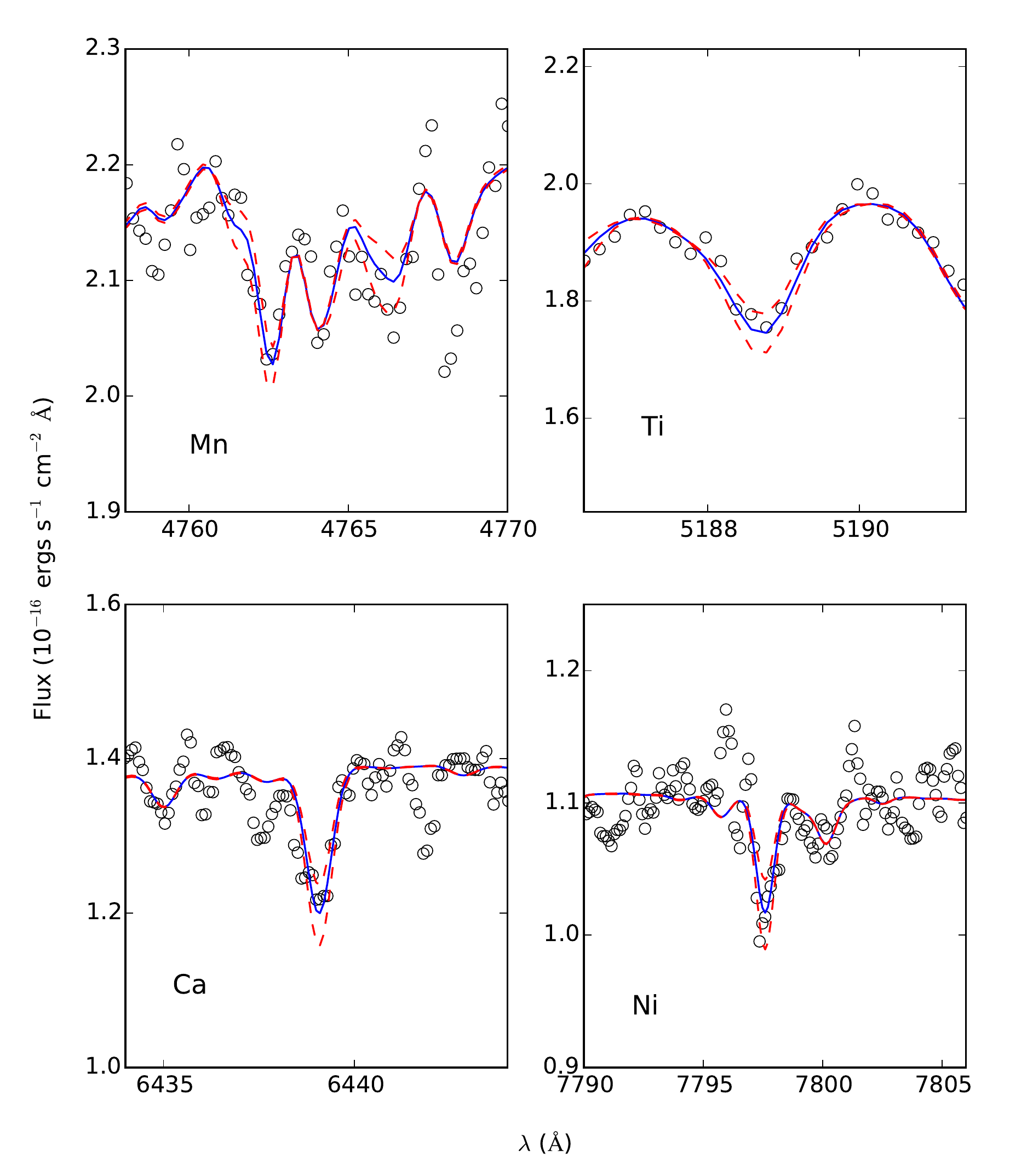}}
      \caption{Example synthesis fits for NGC1313-379. Empty black circles belong to the X-Shooter observations. Blue curve shows the best abundance for the element specified in the subplots. Red dashed curve shows the effect of varying the element in question by $\pm$0.5 dex. }
         \label{Fig:Spec13}
   \end{figure}
   
   \begin{figure}
   \resizebox{\hsize}{!}
            {\includegraphics[scale=0.5]{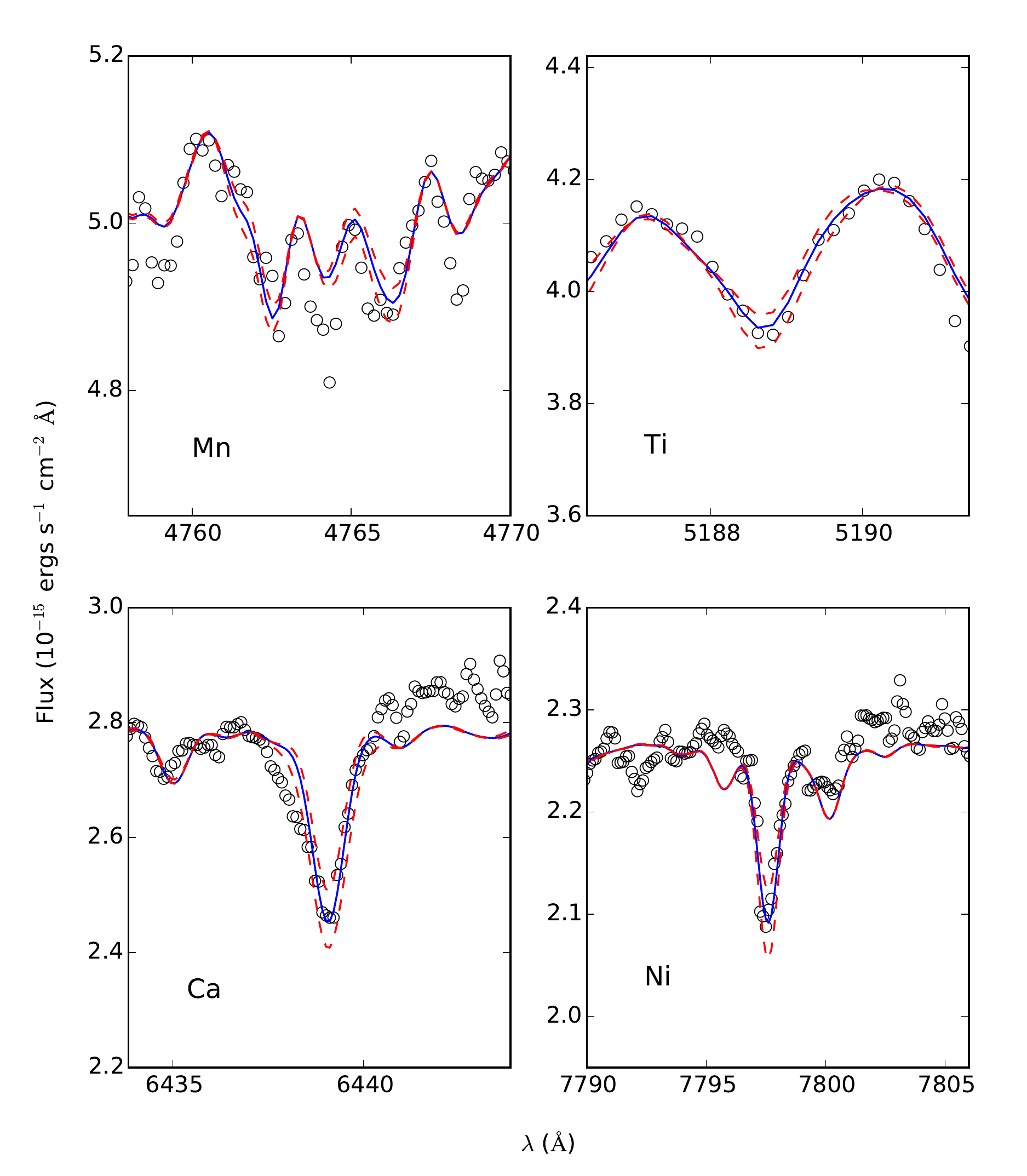}}
      \caption{Example synthesis fits for NGC1705-1. Empty black circles belong to the X-Shooter observations. Blue curve shows the best abundance for the element specified in the subplots. Red dashed curve show the effect of varying the element in question by $\pm$0.5 dex. }
         \label{Fig:Spec17}
   \end{figure}

We present in Table \ref{table:avg} weighted average abundances, their corresponding errors ($\sigma_\mathrm{w}$) 
and the number of bins ($N$) for both NGC1313-379 and NGC1705-1 using the Isochrone-Only method. The weighted errors are computed as

\begin{equation}
\sigma_\mathrm{w} = \sqrt{\frac{1}{\sum w_i}}
\end{equation}

\noindent where the individual weights, $w_i$, are estimated by $w_i = 1/ \sigma_{i}^2$, and $\sigma_{i}$ are the 1-$\sigma$ uncertainties listed in Tables \ref{table:chem379} and \ref{table:chem1}. The standard deviation, $\sigma_\mathrm{STD}$, appears to be more representative of the actual measurement uncertainties. We see that the scatter in the individual measurements is larger than the formal errors based on the $\chi^2$ analysis. For this we turn the $\sigma_\mathrm{STD}$ into errors on the mean abundances by accounting for the number of individual measurements ($N$);

\begin{equation}
\sigma_\mathrm{err} = \frac{\sigma_\mathrm{STD}}{\sqrt{N-1}}
\end{equation}

\noindent We proceed quoting $\sigma_\mathrm{err}$ as the measurement error when applicable. Additionally, we include this error estimate in Table \ref{table:avg}. Figures \ref{Fig:Spec1317}, \ref{Fig:Spec13}, and \ref{Fig:Spec17} show the best model fits for both YMCs.

\subsection{Systematic uncertainties in $T_\mathrm{eff}$} \label{Teff}
\citet{dav13} find that specifically for RSGs, estimating the $T_\mathrm{eff}$ with a combination of their \textit{V-I} colours and model atmosphere colour-$T_\mathrm{eff}$ transformations can lead to substantially underestimated temperature values. In their work \citeauthor{dav13} study the RSG population in both the LMC and SMC and show that the discrepancies between $T_\mathrm{eff}$ estimated from the \textit{VRI} spectral region and those obtained from the \textit{IJHK} regions arise from the complexity of the TiO bands dominating the optical part of the the RSG spectrum.\par
To evaluate the systematic uncertainties originating from the RSG colour-$T_\mathrm{eff}$ transformation, we take the CMD+ISO parameters of NGC1313-379 and manually assign a $T_\mathrm{eff}$ = 4150 K to all supergiants in the cluster. This temperature was chosen based on the results of \citeauthor{dav13}, where all RSGs in both Magellanic Clouds have a uniform temperature of $T_\mathrm{eff}$ = 4150 K. Here we define supergiant as any object with $\log g \leq$ 1.0. This change in the input stellar parameters decreases the measured [Fe/H] only by $\sim$0.04 dex. We note that the mean $T_\mathrm{eff}$ estimated from the colour-$T_\mathrm{eff}$ transformation for NGC1313-379 is $T_\mathrm{eff}$ $\sim$ 4300 K, only 150 K higher than that from \citeauthor{dav13}.\par
We perform the same exercise on NGC1705-1, manually modifying the $T_\mathrm{eff}$ of all supergiants in the input file. In contrast to the decrease observed in NGC1313-379, we measure a slightly higher metallicity (0.04 dex) for NGC1705-1, [Fe/H] $\sim -$0.74. We point out that the original mean $T_\mathrm{eff}$ of the supergiants in NGC1705-1 is $T_\mathrm{eff}$ $\sim$ 3800 K.

\subsection{Sensitivity to input isochrone properties} \label{sens_iso}
The analysis of the two YMCs presented in this paper relies heavily on the use of theoretical isochrones.  The selection of such models is based on assumptions in the age and metallicity of the clusters, mainly values found in the literature. Due to this intrinsic dependence on the age and metallicity of the clusters, we consider the uncertainties involved in the selection of a single isochrone. To explore these uncertainties, we repeat the analysis done on both YMCs and the results are as follows.

 \subsection*{NGC1313-379}
We recalculate the abundances using different choices of isochrones, varying the age and metallicity and the method selection (CMD+Isochrone and Isochrone-Only). In Table \ref{table:iso_sensitivity} we present the results of this parameter study. For NGC1313-379 we find that the differences in the abundances obtained when using CMD+Isochrone and Isochrone-Only methods have a minor effect on most abundances, on the order of $\lesssim$ 0.1 dex, with the exception of [Sc/Fe]. It is important to note that the measurements for this element are based on a single wavelength bin, which makes it difficult to assess the true uncertainties.\par
When the isochrone age is changed by 25 Myrs, using the Isochrone-Only method as a test case, the Fe abundance changes by less than 0.1 dex. Regarding the individual element abundances, we see that the sensitivity of $\alpha$-element abundances to age variations are on the order of 0.1 dex as well. The uncertainties for Fe-peak abundances, depending on the element, are larger than those for $\alpha$-elements with typical differences of <0.2 dex, except for [Sc/Fe]. Once more, for Sc we see that the uncertainties are the highest when changing the age by 25 Myrs.\par
An isochrone change in metallicity of $+$0.15 dex causes the derived [Fe/H] abundance to increase by $\sim$0.03 dex. The impact of such a change in the rest of the elements is $\leq$ 0.1 dex. \par

 \subsection*{NGC1705-1}
We repeat the same steps described for NGC1313-379, but now for NGC1705-1. The results of this sensitivity study for NGC1705-1 are presented in Table \ref{table:iso_sensitivity_1705}. With the exception of [Fe/H[, for those changes involving age and metallicity we see that the uncertainties are greater for NGC1705-1 than those observed in NGC1313-379. We believe this is mainly driven by the young age of NGC1705-1. At ages of several $\times 10^7$ years the models for massive stars in these YMCs are much less certain than low-mass stars in GCs \citep{mas03}.\par
Despite the observed trends when using the CMD+Isochrone method, we average over the values in order to get a sense of the difference and uncertainties between the two approaches. As observed throughout this procedure, the [Fe/H] abundance only changes by $\sim -$0.02 dex when using the CMD+Isochrone in place of the Isochrone-Only method. This change also affects the [Mg/Fe] ratio abundance only slightly, with an increase of $\ll +$0.01 dex. [Ca/Fe], on the other hand increases by as much as $\sim +$0.20 dex. The rest of the abundance ratios are decreased by $\sim$ $-$0.30 dex. 
Changing the age of the input isochrone by $+$25 Myr increases the [Fe/H] ratio by 0.050 dex, a similar change to that observed in NGC1313-379. The $\alpha$-element ratios with respect to Fe changed by $\sim$ 0.1 dex, with $\Delta$[Ca/Fe] = $-$0.168 dex being the highest. The Fe-peak elements, on the other hand, experience changes of $\gtrsim$ 0.3 dex. \par
An isochrone change in metallicity of $+$0.15 dex causes a decrease in the derived [Fe/H] abundance of $\sim -$0.03 dex. For this YMC, the [Mg/Fe] ratio is also negligibly affected by this change, with the abundance ratio decreasing by < 0.1 dex. The abundances estimated for the rest of the $\alpha$-elements (Ca and Ti) change by $\sim$ 0.2-0.3 dex. This same change modifies the measured abundances for Fe-peak elements by $\sim$ 0.4 dex, and in the case of [Ni/Fe] the abundance change is as high as $\sim$ 0.7 dex.  

\begin{table}
\caption{Averaged abundance measurements using the Isochrone-Only method}
\label{table:avg}
 \centering 
\begin{tabular}{ccccc} 
 \hline  \hline \
Element& Weighted Avg&$\sigma_\mathrm{w}$&$\sigma_\mathrm{err}$&$N$\\
\hline
\multicolumn{5}{c}{NGC1313-379} \\
\multicolumn{1}{l}{[Fe/H]}&$-$0.843& 0.014&0.065&6 \\
\multicolumn{1}{l}{[Mg/Fe]}&$+$0.124&0.044&0.305&2\\
\multicolumn{1}{l}{[Ca/Fe]}&$+$0.114&0.008&0.073&6\\
\multicolumn{1}{l}{[Sc/Fe]}&$+$0.350&0.242& -&1\\
\multicolumn{1}{l}{[Ti/Fe]}&$-$0.060&0.053&0.079&3\\
\multicolumn{1}{l}{[Cr/Fe]}&$+$0.479&0.095&0.073&3\\
\multicolumn{1}{l}{[Mn/Fe]}&$-$0.331&0.252& -&1\\
\multicolumn{1}{l}{[Ni/Fe]}&$+$0.456&0.049&0.135&6\\
 \hline 
 \multicolumn{5}{c}{NGC1705-1} \\
\multicolumn{1}{l}{[Fe/H]}&$-$0.775&0.011&0.099&6\\
 \multicolumn{1}{l}{[Mg/Fe]}&$+$0.274&0.009&0.197&2\\
 \multicolumn{1}{l}{[Ca/Fe]}&$+$0.218&0.004&0.285&5\\
 \multicolumn{1}{l}{[Sc/Fe]}&$+$0.192&0.052& -&1\\
 \multicolumn{1}{l}{[Ti/Fe]}&$+$0.462&0.030&0.120&3\\
 \multicolumn{1}{l}{[Cr/Fe]}&$-$0.270&0.084&0.553&2\\
 \multicolumn{1}{l}{[Mn/Fe]}&$-$0.229&0.242& -&1\\
 \multicolumn{1}{l}{[Ni/Fe]}&$+$0.742&0.028&0.485&5\\
 \hline
 \end{tabular}
\end{table}

\begin{table}
\caption{Sensitivity to input isochrone parameters for NGC1313-379.}
\label{table:iso_sensitivity}
 \centering 
\begin{tabular}{cccc} 
 \hline  \hline \
Element& CMD+ISO/&\multicolumn{1}{c}{$\Delta$ t}&{$\Delta$ Z}\\\
&Only ISO&$+$25Myr&$+$0.15 dex \\
\hline
\multicolumn{1}{l}{$\Delta$ [Fe/H]}&$-$0.056&$-$0.086&$+$0.033 \\
\multicolumn{1}{l}{$\Delta$[Mg/Fe]}&$-$0.129&$-$0.040&$+$0.001\\
\multicolumn{1}{l}{$\Delta$[Ca/Fe]}&$-$0.180&$-$0.128&$+$0.043\\
\multicolumn{1}{l}{$\Delta$[Sc/Fe]}&$-$0.254&$-$0.348&$+$0.077\\
\multicolumn{1}{l}{$\Delta$[Ti/Fe]}&$+$0.026&$-$0.090&$+$0.018\\
\multicolumn{1}{l}{$\Delta$[Cr/Fe]}&$+$0.039&$-$0.136&$-$0.065\\
\multicolumn{1}{l}{$\Delta$[Mn/Fe]}&$-$0.067&$-$0.050&$+$0.185\\
\multicolumn{1}{l}{$\Delta$[Ni/Fe]}&$+$0.095&$-$0.151&$+$0.041\\
 \hline 
 \end{tabular}
\end{table}

\begin{table}
\caption{Sensitivity to input isochrone parameters for NGC1705-1.}
\label{table:iso_sensitivity_1705}
 \centering 
\begin{tabular}{cccc} 
 \hline  \hline \
Element& CMD+ISO/&\multicolumn{1}{c}{$\Delta$ t}&{$\Delta$ Z}\\\
&Only ISO&$+$25Myr&$+$0.15 dex \\
\hline
\multicolumn{1}{l}{$\Delta$ [Fe/H]}&$-$0.016&$+$0.050&$-$0.025 \\
\multicolumn{1}{l}{$\Delta$[Mg/Fe]}&$+$0.001&$+$0.079&$-$0.087\\
\multicolumn{1}{l}{$\Delta$[Ca/Fe]}&$+$0.231&$-$0.168&$+$0.309\\
\multicolumn{1}{l}{$\Delta$[Sc/Fe]}&$-$0.313&$-$0.288&$+$0.363\\
\multicolumn{1}{l}{$\Delta$[Ti/Fe]}&$-$0.371&$-$0.004&$+$0.217\\
\multicolumn{1}{l}{$\Delta$[Cr/Fe]}&$-$0.023&$+$0.363&$+$0.407\\
\multicolumn{1}{l}{$\Delta$[Mn/Fe]}&$-$0.304&$+$0.594&$+$0.498\\
\multicolumn{1}{l}{$\Delta$[Ni/Fe]}&$+$0.028&$+$0.245&$+$0.663\\
 \hline 
 \end{tabular}
\end{table}

\subsection{Balmer emission lines} \label{bal_em}
The spectroscopic observations of both YMCs display strong Balmer emission lines, especially in H$\alpha$ (Figure ~\ref{Fig:balmer}). The Balmer emission line in NGC1705-1 has previously been observed by \citet{mel85}. In their work \citeauthor{mel85} estimate a velocity dispersion of $\sigma_{1D}$ $\sim$ 130 km s$^{-1}$, which they point out is too high to be produced by gas within the YMC. One possibility for the origin of these emission lines, other than gas, is Be stars. These are B-type non-supergiant stars, characterised by rotational velocities of several hundreds of km s$^{-1}$ \citep{mar82} that show strong H$\alpha$ emission lines \citep{tow04}. The broad Balmer emissions observed in both YMCs hint at the presence of Be stars. Studies have found an enhancement of Be stars in young clusters \citep[<100 Myr,][]{mcs05,wis06,mat08}. At $\sim$12 and 56 Myr, NGC1313-379 and NGC1705-1 are found within the range of ages where high fractions of Be stars are expected.

\section{Discussion}\label{discussion}
In this section we discuss our results and compare our measurements to those observed in similar environments to NGC1313 and NGC1705. \par
NGC1313 is a late-type barred spiral with morphological type SB(s)d. Using H II regions, \citet{wal97} estimated a metallicity of 12$+$log O/H $\approx$ 8.4 $\pm$ 0.1 or [O/H] $=$ $-$0.43 $\pm$ 0.1 \citep{gre98}, similar to that of young stellar populations and H II regions in the LMC. \par

NGC1705, on the other hand, is a late-type galaxy classified as a blue compact dwarf (BCD). NGC1705 shows significant recent star formation activity, clear evidence of galactic winds \citep{meu92} and a metallicity similar to that measured in the SMC \citep{lee04}. BCDs are low metallicity systems exhibiting on-going or recent bursts of star formation, but otherwise appear to have had roughly continuous star formation histories in their past \citep{izo99,tol09}. NGC1705 is no exception and has been observed to host high gas content and have experienced recent star formation activity. Galactic winds appear to be the most viable process to explain the coexistence of high star formation rates and low metal abundances in these galaxies \citep{mat85,mar94,car95,rom06}.

\subsection{Fe} \label{dis_Fe}
 \subsection*{NGC1313-379}
To the best of our knowledge no previous determination of stellar [Fe/H] has been published for NGC1313. We measure an Fe abundance of $-$0.84 $\pm$ 0.07, slightly lower ($\sim$ 0.3 dex) than what would have been expected based on previous studies of H II regions in NGC1313. Our inferred metallicity is also lower than the metallicity ranges that \citet{col12} find in three young star clusters in the LMC ($-$0.57 < [Fe/H] < $+$0.03), a galaxy which may be expected to have a comparable metallicity to that of NGC1313, based on its luminosity. \par
\citet{sil12} studied the star formation history of NGC1313 mainly focusing on three regions: the northern, southern,  and southwest fields. The observations of \citeauthor{sil12} for the southwest region as a whole (the region which includes NGC1313-379) showed lower levels of star formation when compared to those obtained for the northern and southern fields. A possible explanation for the relatively low [Fe/H] abundance measurement in NGC1313-379 is a lower star formation level relative to the rest of the galaxy. In their work \citeauthor{sil12} propose a scenario where NGC1313 interacted with a satellite companion, an event that triggered an increased star formation rate in the southwest region forming the YMC in question. \par
To verify our metallicity measurement, we attempted to estimate [Fe/H] using the NIR X-Shooter observations; however the S/N$\sim$10 is too low for this type of analysis. Going back to the UVB X-Shooter observations, in Figure ~\ref{Fig:Fe} we show that varying the Fe abundance by $\pm$0.5 dex, from the measured [Fe/H] = $-$0.84,  proves to be a mismatch to the observations. It is important to note that given the model assumptions for NGC1313-379 described in Section ~\ref{ind_cl} our [Fe/H] measurement seems robust; however, problems with the CMD assumptions are still possible, although unlikely for this specific cluster considering the good agreement between the CMD+Isochrone and Isochrone-Only approaches. 

 \subsection*{NGC1705-1}
For NGC1705-1 we measure an Fe abundance of $-$0.78 $\pm$ 0.10. To verify this measurement, we estimate the [Fe/H] for this YMC using the NIR portion of the X-Shooter observations. In this wavelength range, as explained in Section ~\ref{intro}, the spectrum is less affected by uncertainties in the CMD modelling assumptions. This is because the NIR stellar continuum at these young ages is  entirely dominated by the flux from RSGs \citep{ori00}. Without any tailored wavelength windows, we calculate the overall metallicity and [Fe/H] abundance in 200 $\AA$ bins, excluding those regions where telluric absorption is the strongest. For this test we also use the Isochrone-Only approach and extract the stellar parameters from an isochrone of log(age)=7.1 and metallicity Z=0.008. The wavelength bins, [Fe/H] abundances and their corresponding 1-$\sigma$ uncertainties are presented in Table ~\ref{table:nir1705}. Taking into consideration only those bins with reduced $\chi^2$ < 1.5 (all, except the 12200.0-12400.0 $\AA$ bin), this results in a weighted average [Fe/H] = $-$0.73 $\pm$ 0.07. The [Fe/H] abundance measured using the UVB and VIS observations is within the errors of the NIR [Fe/H] abundance. From this test we conclude that the [Fe/H] is reliably measured. To further confirm the robustness of this measurement, we show in Figure ~\ref{Fig:Fe} example synthesis fits for an Fe line and the effect of varying the Fe abundance by $\pm$ 0.5 dex.\par

Since the metallicity of NGC1705 has been previously estimated to be similar to that of the SMC we now compare our [Fe/H] abundance to that measured in the SMC. \citet{hill99} studied the brightest young cluster in the SMC, NGC330, and measured an abundance of [Fe/H]= $-$0.82 $\pm$ 0.11, along with an SMC [Fe/H] = $-$0.69 $\pm$ 0.10 for field stars. Our measured [Fe/H] abundance is comparable and within 1-$\sigma$ error to that obtained by \citet{hill99} for the SMC young cluster NGC330. In this work we see that the stellar component in NGC1705-1 has a metallicity similar to the young stellar components in the SMC. In a separate study \cite{alo05} looked into the Interstellar Medium (ISM) of NGC1705 using STIS and FUSE observations and measured a total (neutral + ionised) [Fe/H] = $-$0.86 $\pm$ 0.03, similar to our measured [Fe/H] for NGC1705-1. Furthermore, NGC1705-1 has also been compared to NGC1569-B as they both display similar properties; for example, CMD morphology, host galaxies with strong galactic winds, and active star formation \citep{lar08,ani03, wal91}. \citeauthor{lar08} studied NGC1569-B and found an iron abundance of [Fe/H] = $-$0.63 $\pm$ 0.08, similar to that of field stars in the SMC \citep{hill99}, and slightly higher ($\sim$0.15 dex) than what we have measured for NGC1705-1. 

\begin{table}
\caption{[Fe/H] measurements for NGC1705-1 using NIR X-Shooter observations}
\label{table:nir1705}
 \centering 
\begin{tabular}{cccc} 
 \hline  \hline \
Element& $Wavelength\, [$\AA$]$&Abundance &Error\\
  \hline\
{[Fe/H]}& 10200.0-10400.0& $-$0.706 & 0.121 \\
& 10400.0-10600.0& $-$0.745 &0.091 \\
& 10600.0-10800.0& $-$0.825 &0.121 \\
& 12200.0-12400.0& $-$0.614 &0.090 \\
& 12400.0-12600.0& $-$0.695 &0.141 \\
& 12600.0-12800.0& $-$0.420 &0.215 \\
 \hline 
 \end{tabular}
\end{table}

\subsection{$\alpha$-elements} \label{dis_1313_alpha}
Figure \ref{Fig:abun1313} shows the [Mg/Fe], [Ca/Fe], and [Ti/Fe] abundances calculated as part of this work and compares them to abundance trends observed in the MW, LMC, and M31. 
The individual $\alpha$ abundances for both NGC1313-379 and NGC1705-1 are displayed in Figure \ref{Fig:abun1313} as red triangles and squares, correspondingly. 

 \subsection*{NGC1313-379}
\begin{figure}
   \resizebox{\hsize}{!}
            {\includegraphics[]{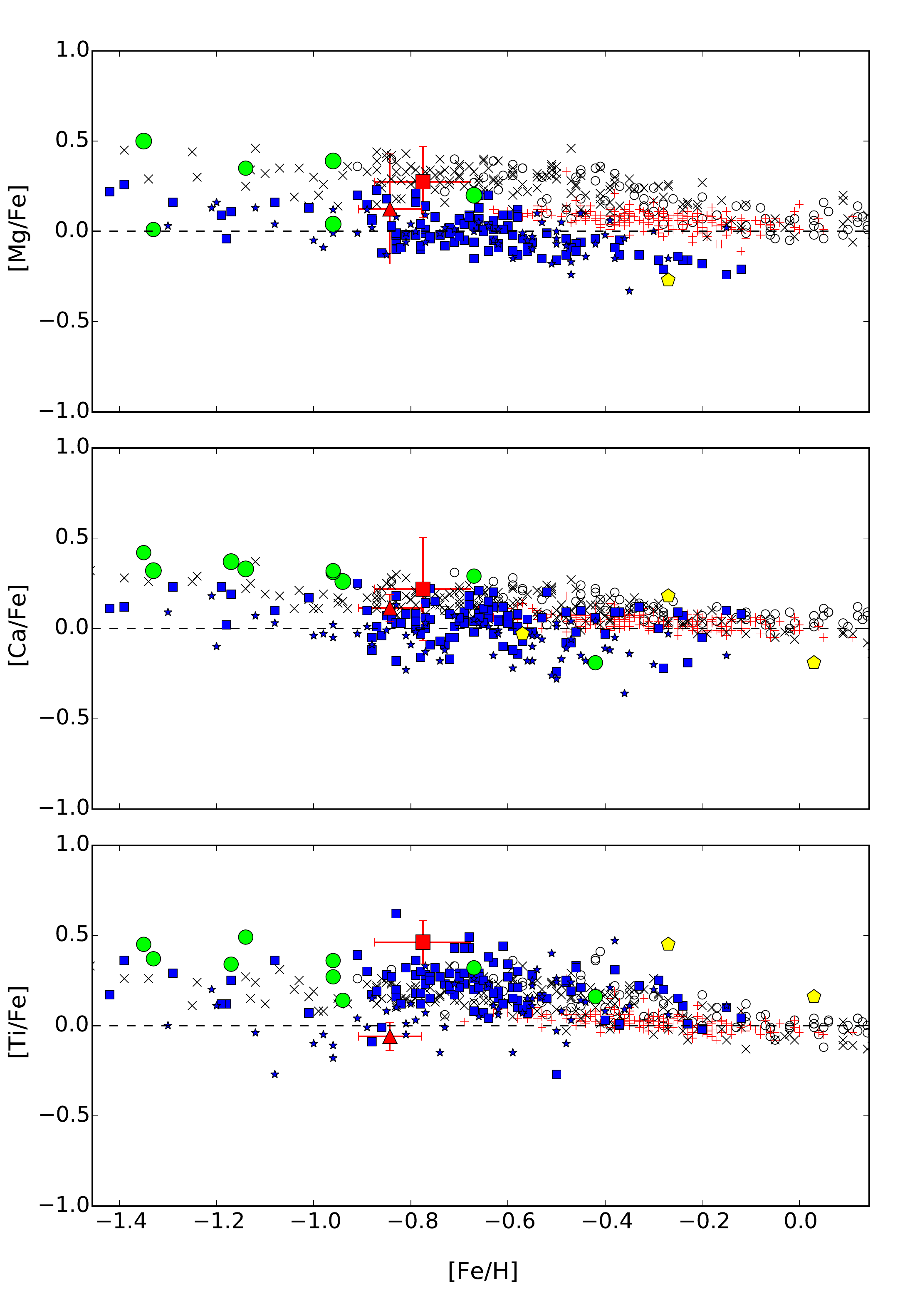}}
      \caption{Top: [Mg/Fe] vs. [Fe/H]. Middle: [Ca/Fe] vs. [Fe/H]. Bottom: [Ti/Fe] vs. [Fe/H]. Legend: red triangles show the alpha abundances estimated for NGC1313-379 (this work); red squares display the measurements for NGC1705-1; red crosses and black exes present MW disc abundances from \citet{red03,red06}, respectively; black open circles are MW disc abundances from \citet{ben05}; blue squares and stars belong to LMC bar and inner disc abundances presented by \citet{van13}; yellow pentagons belong to abundances for LMC young star clusters by \citet{col12}; green circles are abundance measurements from GCs in both the MW and M31 \citep{col09,cam09}.}
         \label{Fig:abun1313}
   \end{figure}

We compare our individual NGC1313-379 abundances to those measured in what is expected to be a similar environment, the LMC. We measure a [Mg/Fe] abundance of $+$0.12 $\pm$ 0.31, within the range of abundances measured in the LMC for a metallicity of [Fe/H] =$-$0.84 (see Figure \ref{Fig:abun1313}). However, it is worth mentioning that metallicities around [Fe/H] $\sim -$ 0.84 dex are found in relatively old stars in the LMC (> 3 Gyr, \citealt{van13}). For the [Ca/Fe] ratio we measure an abundance of $+$0.11 $\pm$ 0.07, comparable to the abundances measured in the LMC. Measurements of Ti abundances in the LMC seem to display a noticeable scatter (bottom plot in Figure \ref{Fig:abun1313}) and our measurement of [Ti/Fe] = $-$0.06 $\pm$ 0.08 appears to be on the lower part of the envelope of values observed in the LMC. All three measurements of $\alpha$ elements are consistent with the Solar-scaled values within the uncertainties. This behaviour is expected for young stellar populations and has previously been observed in other environments, like the MW, LMC, SMC and M31.\par
Averaging our $\alpha$-abundances (Ti, Ca and Mg) for NGC1313-379, we measure a close-to-solar [$\alpha$/Fe] =$+$0.06 $\pm$ 0.11 or [$\alpha$/H] = $-$0.78 $\pm$ 0.11. As O is an $\alpha$ element, we compare our [$\alpha$/H] results with the work of \citet{wal97} who find an average [O/H] $=$ $-$0.43 $\pm$ 0.10. From this comparison we see that our measurements of $\alpha$ abundance are lower by $\sim$ 0.35 dex. However, using ([O II] + [O III])/H$\beta$ to derive [O/H] abundances, \citeauthor{wal97} measured [O/H] $\sim$ $-$0.73 and $-$0.43 for two H II regions in close proximity to NGC1313-379, the former agreeing with our inferred [$\alpha$/H]. We note that these two H II regions appear to be part of the same complex as that of NGC1313-379. \par

 \subsection*{NGC1705-1}
We measure an abundance ratio of [Mg/Fe] = $+$ 0.27 $\pm$ 0.20 for NGC1705-1. \citet{hill97} studied field stars in the SMC and measured [Mg/Fe] abundance ratios in the range $-$0.01 < [Mg/Fe] < $+$ 0.29. Comparing our [Mg/Fe] measurement to the findings of \citet{hill97} we see that our abundance ratio is within those measured in field stars in the SMC. 
Additionally, we infer super-solar [Ca/Fe] = $+$0.22 $\pm$ 0.29 and [Ti/Fe] = $+$ 0.46 $\pm$ 0.12 for NGC1705-1. In general we observe a super-solar trend in all three measured $\alpha$ elements, which is unusual for a young population. \par
\citet{lee04} measured the oxygen abundance for NGC1705 using long-slit spectroscopic observations of 16 H II regions. They reported a mean oxygen abundance of 12$+$log(O/H) $=$ 8.21 $\pm$ 0.05, corresponding to [O/H] $= -$0.63 \citep{gre98}. The abundance measured by \citeauthor{lee04} is at the middle of the range of values reported in different studies \citep{meu92,sto94,sto95,hec98} varying from 12$+$log(O/H) $=$8.0 to 8.5 or  [O/H] $= -$0.83 to $-$0.33 \citep{gre98}. The same study finds an absence of radial gradient in the oxygen metallicity.\par
A recent study by \citet{ann15} using multi-object spectroscopy and narrow-band imaging of [OIII] inferred oxygen metallicities of PNe and H II regions distributed throughout the galaxy. They detected a negative radial gradient in the oxygen metallicity of $-$0.24 $\pm$ 0.08 dex kpc$^{-1}$. Additionally, they found an average oxygen abundance of 12$+$log(O/H) $=$ 7.96 $\pm$ 0.04 or [O/H] $= -$0.87, $\sim$0.25 dex lower than the value reported by \citet{lee04}. This average is calculated including only H II regions located within 0.4 kpc of the centre of the galaxy.\par
Comparing our average $\alpha$-abundance  (Ti, Ca and Mg)  for NGC1705-1 with the work of \citet{lee04}, we see that our measurement of  [$\alpha$/H] = $-$0.46 $\pm$ 0.12 is slightly higher ($\sim$ 0.15 dex) than that measured by \citeauthor{lee04} in H II regions. In contrast, our average $\alpha$-abundance is $\sim$0.4 dex higher than what \citeauthor{ann15} report. We note that the oxygen measurements from \citeauthor{lee04} and \citeauthor{ann15} mainly characterise and describe the properties of the ionised gas in NGC1705, whereas our work studies the stellar component of NGC1705-1.\par
With the presence of strong galactic winds, NGC1705 has become a target of interest when studying and testing chemical evolution models. Assuming  star formation rates (SFRs) inferred by \citet{ani03}, NGC1705 models by \citet{rom06} predict super-solar [O/Fe] abundance ratios, $+$0.2 < [O/Fe] < $+$0.5 for [Fe/H] $\sim$ $-$0.80, agreeing with our measured [$\alpha$/Fe] = $+$0.32. The observed super-solar [$\alpha$/Fe] ratios were possibly produced by the star formation event observed by \citet{ani03} $\sim$3-35 Myrs before the formation of NGC1705-1. Such a period would have allowed Type II SNe to generate $\alpha$-elements and mix them with pre-existing gas before the YMC formed. 
Similar to NGC1705-1, \citet{lar08} measured a super-solar [$\alpha$/Fe] ratio for NGC1569-B of $+$0.31 $\pm$ 0.09. They also find an oxygen abundance, [O/H], approximately a factor of two higher than derived from H II studies. Again, a possible explanation for this enhancement in both NGC1569-B and NGC1705-1 is that the YMCs formed while the ISM was mainly enriched by Type II SNe products, with genuine differences between abundances in H II regions and the stellar component. However, one cannot rule out the possibility of systematic errors in the method used to measure these $\alpha$ abundances (e.g. in the modelling of the stellar spectra). \par

 \subsection*{}
In summary, we find that all [(Mg,Ca,Ti)/Fe] element ratios in NGC1313-379 are solar (within $\sim$0.1 dex), and super-solar in NGC1705-1. For both YMCs, the comparison of our Mg, Ca and Ti measurements to O abundances by other authors show a consistent picture. From this work and in the context of chemical evolution, NGC1313 resembles several well-studied galaxies with continuous star formation histories such as MW and LMC. NGC1705-1, on the other hand, is more $\alpha$-enhanced as predicted by the chemical evolution models of \citet{rom06} where these abundances may originate from a bursty star formation history.

\subsection{Fe-peak elements} \label{dis_1313_fepeak}
Fe-peak elements are generally produced in SNIa, but the uniqueness of the source for these elements is still debatable. In general, different Fe-peak elements behave differently, especially Mn \citep{mcw97, van13}. \par

 \subsection*{NGC1313-379}
It has been suggested \citep{nis00,red03,red06, van13} that Sc follows a similar pattern as Ca and Ti, decreasing with increasing metallicity. Our measurements for the [Sc/Fe] ratio, $+$ 0.35 $\pm$ 0.24 suggest a mild enhancement in NGC1313-379 when compared to the average [Sc/Fe] abundances in the LMC and other galaxies in the Local Group. \par
[Cr/Fe] abundance ratios in the MW and the LMC overlap each other with abundance ratios from sub-solar to solar  \citep[including individual stars and GCs,][]{joh06,gra91}. We observe an above solar abundance ratio for [Cr/Fe] of $+$0.48 $\pm$ 0.07, with an enhancement higher than what we measure for [Sc/Fe].\par
Mn, on the other hand, has been found to be depleted in metal-poor stars in the MW \citep{mcw97,red06}, but gradually increases to solar values at higher metallicities. It is worth mentioning that the stellar process creating Mn is still uncertain \citep{tim95,she03}. We infer [Mn/Fe] = $-$0.33 $\pm$ 0.25 following the trend observed in environments like the MW at the respective metallicity of [Fe/H] = $-$0.84 dex. \par
For Ni we measure a slightly enhanced [Ni/Fe] = $+$0.46 $\pm$ 0.14. It should be noted that the 7700.0-7800.0 $\AA$ bin yields systematically higher [Ni/Fe] abundance ratios with respect to the rest of the bins. Excluding this bin when averaging our Ni abundances leads to [Ni/Fe] = $+$0.31. This lower [Ni/Fe] ratio is in better agreement with that observed in LMC and MW environments. 

 \subsection*{NGC1705-1}
We measure [Sc/Fe] = $+$0.19 $\pm$ 0.05 for NGC1705-1, which is on the high side of the envelope of observed abundances in both the MW and LMC. In NGC1705-1 we find that the [Cr/Fe] abundance is below solar and agrees well with that observed in our galaxy and the LMC. Our measured [Cr/Fe] = $-$0.27 $\pm$ 0.55 falls within the scatter of the measurements, with relatively large error bars.\par 
We measure [Mn/Fe] = $-$0.23 $\pm$ 0.24, similar to abundances seen in the MW for metallicities of $\sim -$ 0.78 dex. We note that the Mn measurement is based on a single bin with relatively weak lines (See Figure ~\ref{Fig:Spec17}). \par
We infer an enhanced [Ni/Fe] = $+$0.74 $\pm$ 0.49 for this YMC. As pointed out for NGC1313-379, we also measure a consistently higher Ni abundance for the 7700.0-7800.0 $\AA$ bin. Omitting this bin when averaging our Ni abundances leads to a remarkably lower [Ni/Fe] = $+$0.25. We know that Ni is produced in massive stars and in SN Ia \citep{ww95, tim95}. If the enhancements in [Ni/Fe] ratios are genuine, high Ni abundances in this work could indicate that the relative contribution of massive stars and SN Ia to the total Ni production is different for MW, M31 and LMC compared to NGC1705.\par 

 \subsection*{}
Overall, we find that a comparison between the Fe-peak abundances observed in this work and those measured in other environments (MW, LMC and SMC) display a consistent picture for most elements following previously observed trends within their errors. We point out that in general our measurements of these elements are relatively uncertain due to the quality of the observations and would recommend either higher-resolution or higher-S/N data in future work to measure these elements with a higher degree of certainty.

   \begin{figure}
   \resizebox{\hsize}{!}
            {\includegraphics[]{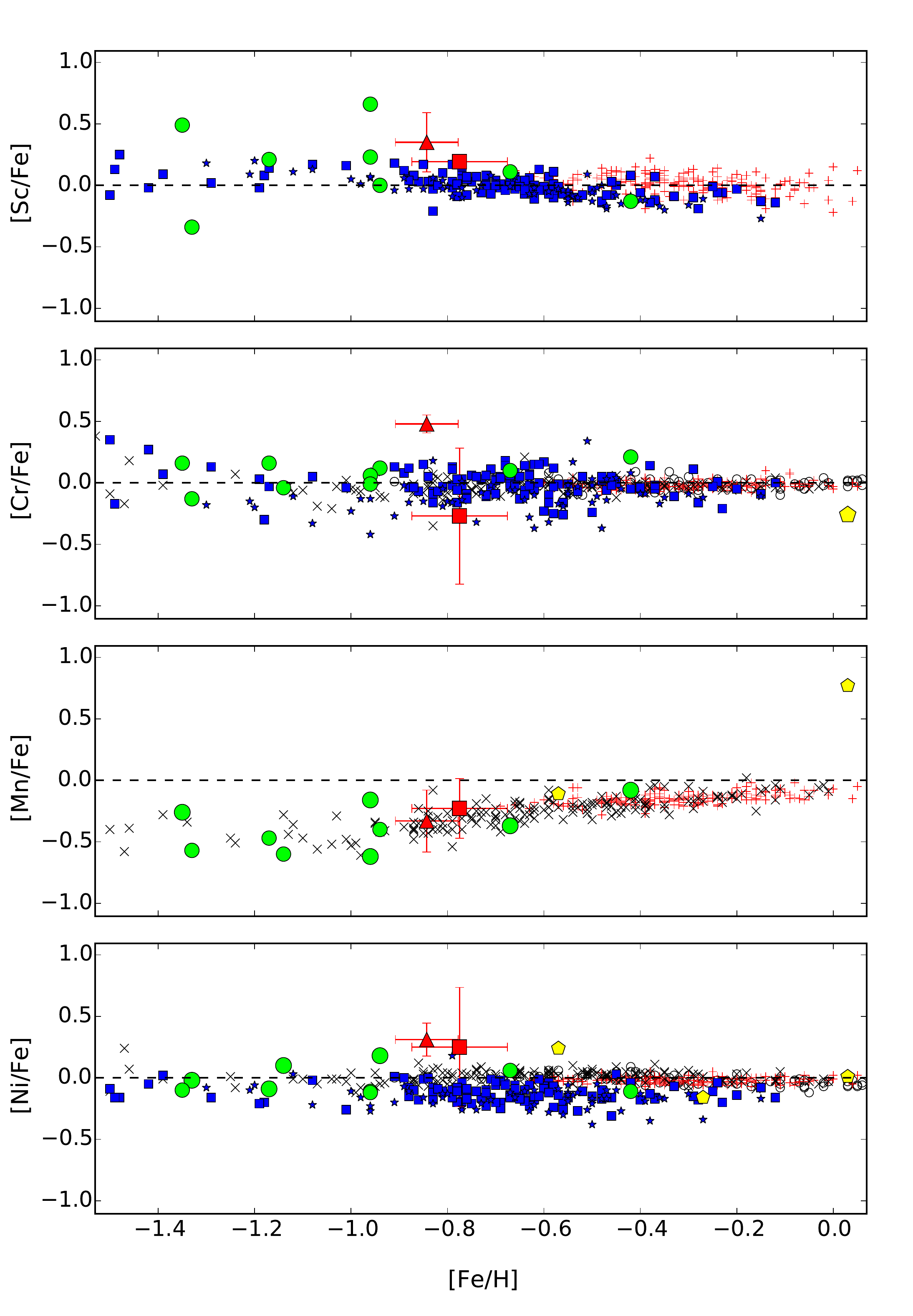}}
      \caption{Top: [Sc/Fe] vs. [Fe/H]. Second row: [Cr/Fe] vs. [Fe/H]. Third row: [Mn/Fe] vs. [Fe/H]. Bottom: [Ni/Fe] vs. [Fe/H]. Symbols as in Figure \ref{Fig:abun1313}.}
         \label{Fig:fepeak1313}
   \end{figure}

\subsection{Multiple populations} \label{MPs}
Studies have shown that, generally speaking, the [Mg/Fe] ratios of field stars, either in the MW, LMC or even in M31, behave similarly to other $\alpha$ elements such as Ca and Ti \citep{ben05,col09, gon11}. In contrast to field stars, studies suggest that intra-cluster Mg variations with respect to other $\alpha$-elements are present in several GCs in the MW \citep{she96,kra97,gra04}. These same variations in Mg were detected as lower [Mg/Fe] abundances when compared to [Ca/Fe] and [Ti/Fe] ratios obtained from integrated-light studies of GCs \citep{mc08,col09,lar14}. The [Mg/Fe] ratios we measure for NGC1313-379 and NGC1705-1, $+$0.12 and $+$0.27 respectively, do not appear to be remarkably lower than any of the other $\alpha$-elements studied as part of this work, hinting at the absence of intra-cluster variations in Mg. The resulting absence of intra-cluster variations in NGC1705-1 in this work is in agreement with the differential analysis performed by \citet{cab16} on NGC1705-1, where they also find a lack of evidence for extreme intra-cluster variations in Al.\par

\section{Conclusions}\label{Conclusions}
In this work we carry out the first detailed abundance and metallicity analysis for two extragalactic YMCs outside of the Local Group, NGC1313-379 and NGC1705-1. The analysis is done following the method and code developed by L12, using a combination of intermediate-resolution integrated-light observations taken with the X-Shooter spectrograph on the ESO Very Large Telescope, color-magnitude diagrams from Hubble Space Telescope and theoretical isochrones from PARSEC.  The main results obtained in this work are summarised as follows:
   \begin{enumerate}
      \item We obtain the first measurements of metallicity in the extragalactic YMCs NGC1313-379 and NGC1705-1 with values of [Fe/H] = $-$0.84 $\pm$ 0.07 and [Fe/H] = $-$0.78 $\pm$ 0.10, respectively.
      \item We obtain the first detailed chemical abundances of Mg, Ca, Ti, Sc, Cr, Mn and Ni for both NGC1313-379 and NGC1705-1.
      \item We find a lack of depletion in the [Mg/Fe] ratios relative to other $\alpha$ elements (Ca and Ti), hinting at the absence of intra-cluster variations in Mg for both YMCs.
      \item We observe strong Balmer emission lines in both YMCs and attribute these phenomena to the presence of Be stars expected in young clusters with ages < 100 Myr.
      \item We measure [$\alpha$/H] = $-$0.78 $\pm$ 0.11 for NGC1313-379 which agrees with the [O/H] $\sim$ $-$0.73  abundance measured by \citet{wal97} for an H II region located in the same complex as NGC1313-379.
      \item The observed super-solar [$\alpha$/Fe] abundance ratio in NGC1705-1 is comparable to $\alpha$-element measurements from a YMC in NGC1569, a galaxy with similar properties as NGC1705 (i.e. strong wind activity). These super-solar [$\alpha$/Fe] ratios were also predicted in NGC1705 by \citet{rom06} for metallicities of [Fe/H] $\sim$ $-$0.80. 
      \item In general Fe-peak elements follow previously observed trends in well studied environments (MW and LMC); however to reduce the uncertainties in the measurements of these elements we recommend higher-resolution observations or higher-S/N data.
         \end{enumerate}
In future work, for those cases where there are no empirical CMDs available, we envision the use of the Isochrone-Only method. In this method the stellar parameters are extracted solely from theoretical models. From this study we have learned that the reduced $\chi^{2}$ obtained when fitting the different abundances can be used as an accuracy indicator for the selection of the optimal isochrone. \par
Having demonstrated that detailed abundance analysis of YMCs outside of the Local Group can be carried out using the same basic methodology that has previously been applied to old GCs, we look forward to combining YMC abundance studies with those of GCs in order to observe the galaxy evolution through a wider window in time. We believe the study of integrated-light observations of stellar clusters continues to be a strong and promising tool to acquire information about the chemical enrichment histories of the host galaxy. 

\begin{acknowledgements}
We would like to thank A. Gonneau, Y.-P. Chen and M. Dries for their help and guidance during the X-Shooter reduction process. Special thanks to our referee, B. Davies, for his careful reading of the manuscript and useful comments which helped improve the quality of this paper. This research has made use of the NASA/IPAC Extragalactic Database (NED), which is operated by the Jet Propulsion Laboratory, California Institute of Technology, under contract with the National Aeronautics and Space Administration.
\end{acknowledgements}

\Online

\begin{appendix} 
\section{Modelled spectra}
Here we present a complete set of modelled spectra which includes our final measured abundances and covers data from UVB and VIS X-Shooter observations for both NGC1313-379 and NGC1705-1. The line IDs are extracted from the linelists found at the Castelli website \footnote{http://wwwuser.oats.inaf.it/castelli/}.

\begin{figure*}
\centering
\includegraphics[angle=270,scale=0.55]{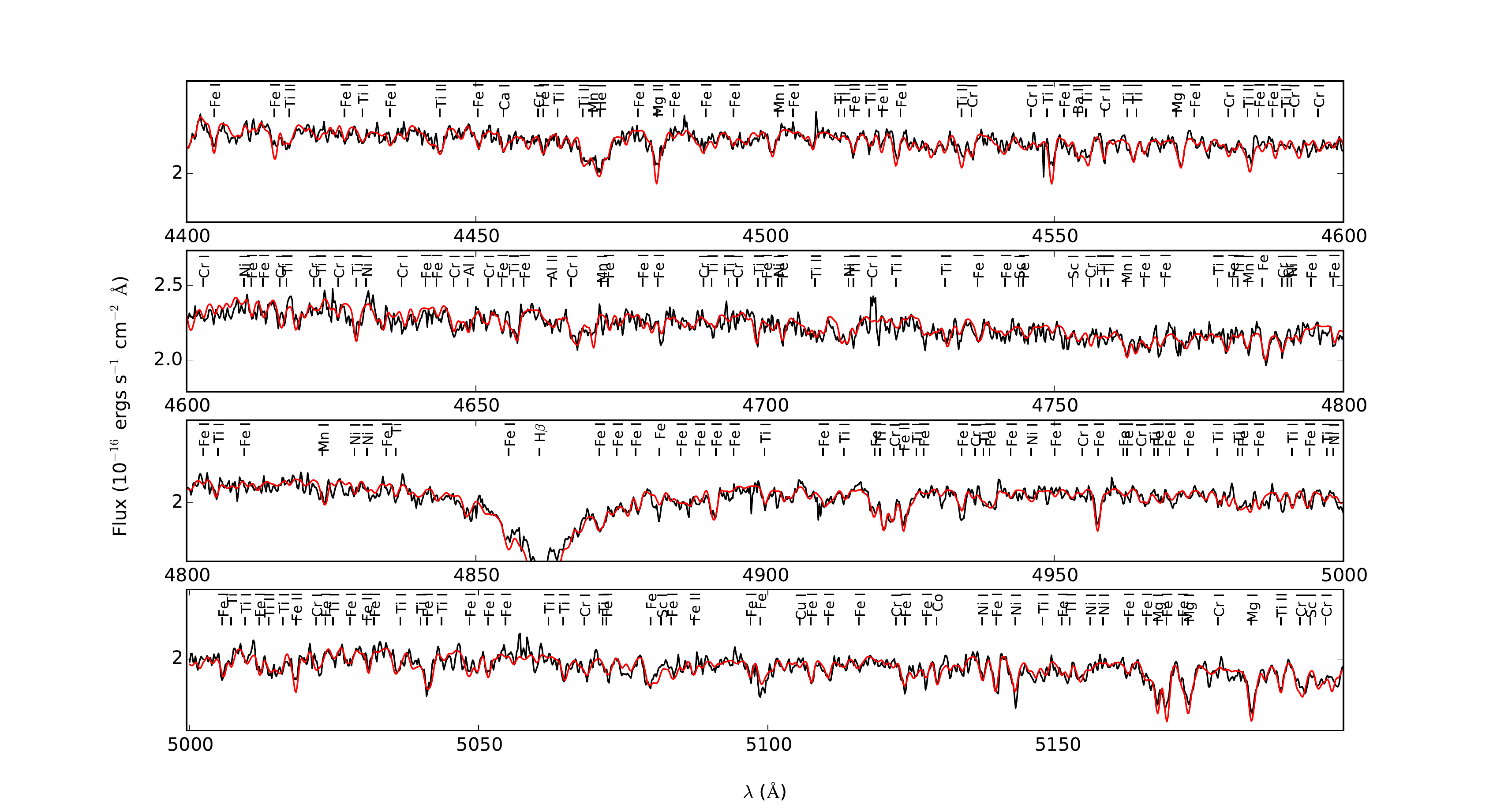}
\caption{Example synthesis fits for NGC1313-379 for different wavelength coverage in the UVB X-Shooter. Black lines show the observations and in red we show the best-fitting model spectra. These models use the final abundances obtained as part of this work. }
\label{appfig1}
\end{figure*}

\begin{figure*}
\centering
\includegraphics[angle=270,scale=0.55]{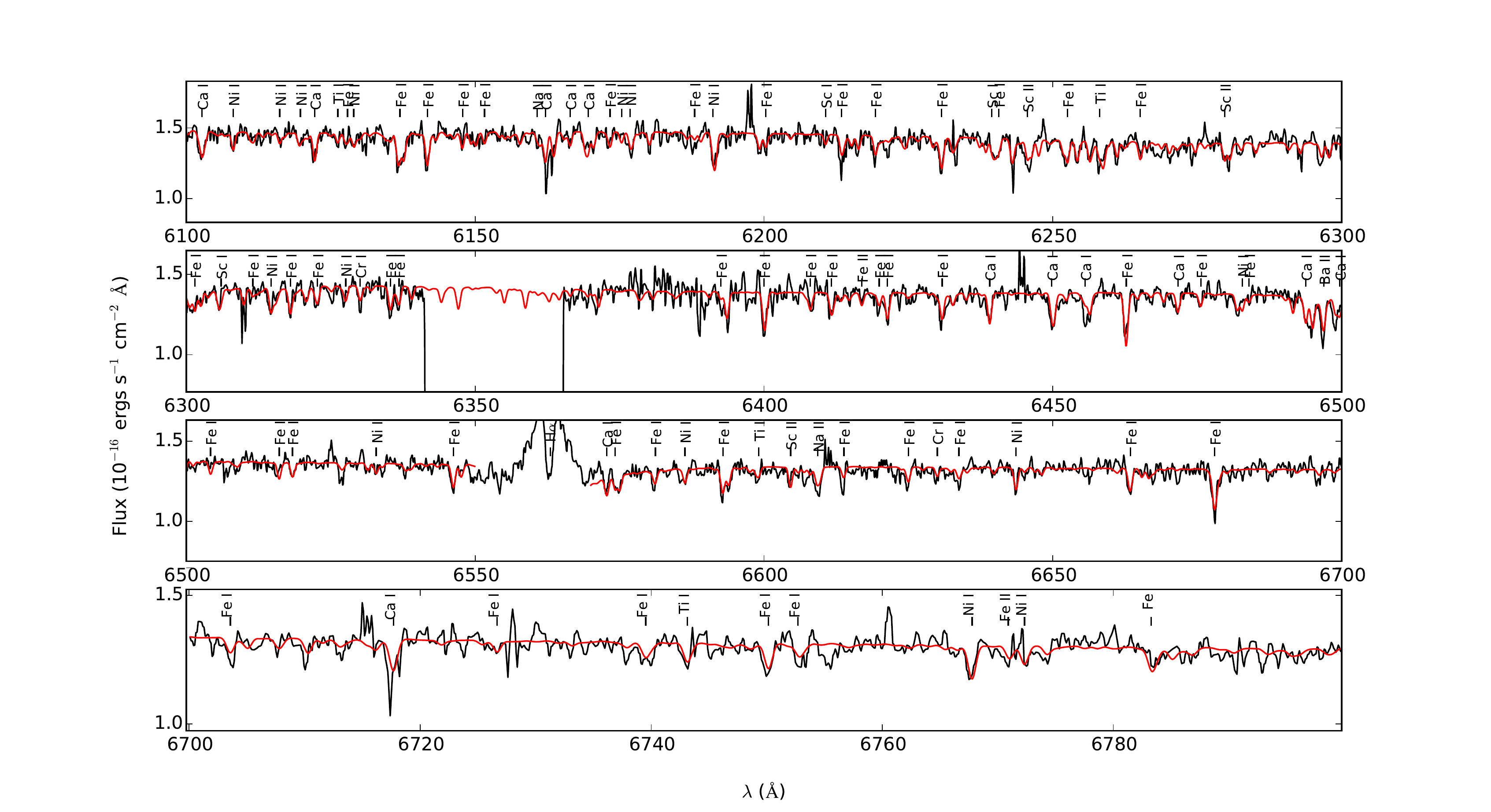}
\caption{Example synthesis fits for NGC1313-379 for different wavelength coverage in the VIS X-Shooter. Here we exclude any telluric contaminated wavelength window. Black lines show the observations and in red we show the best-fitting model spectra. These models use the final abundances obtained as part of this work.}
\label{appfig2}
\end{figure*}

\begin{figure*}
\centering
\includegraphics[angle=270,scale=0.55]{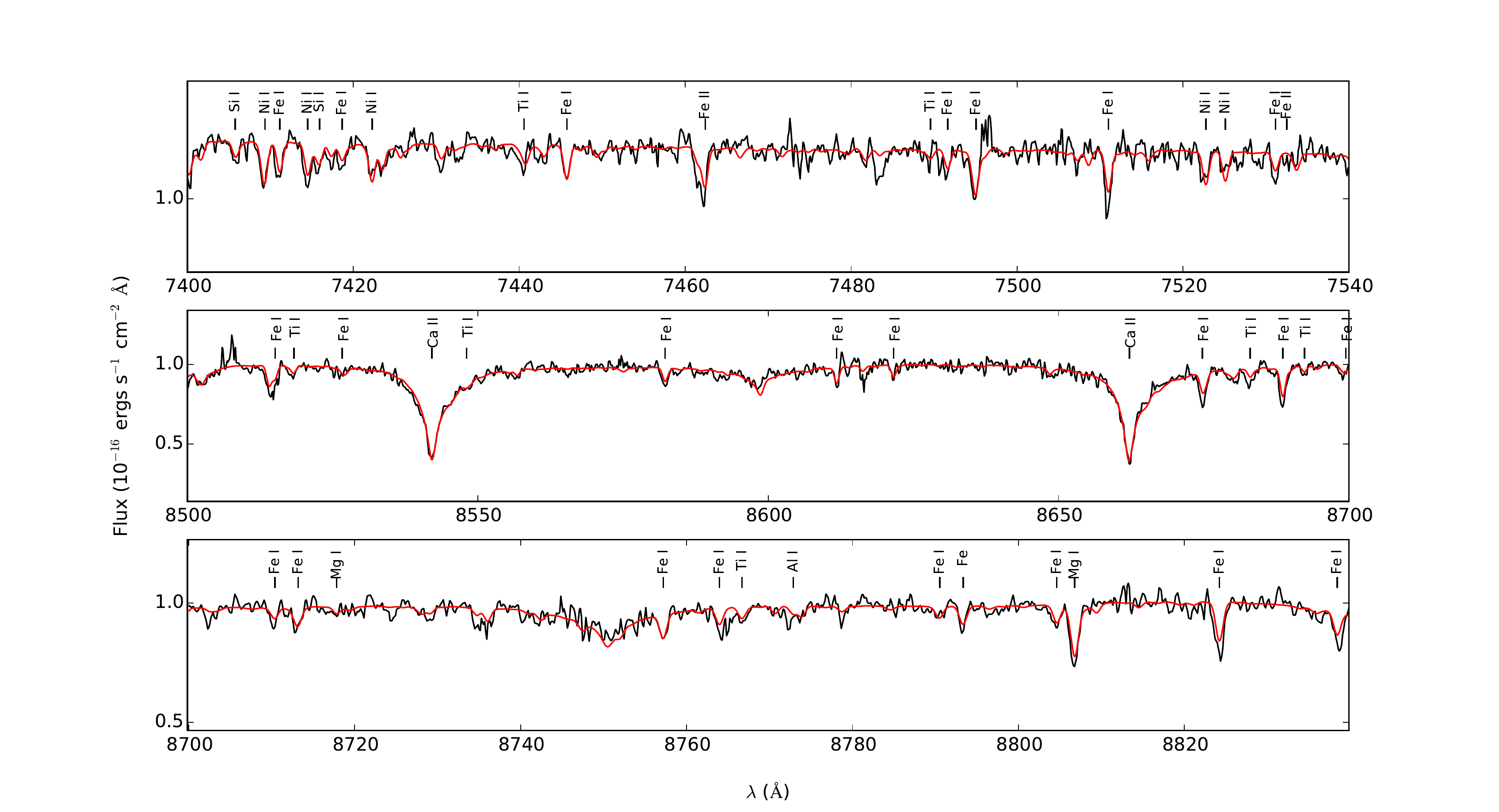}
\caption{Example synthesis fits for NGC1313-379 for different wavelength coverage in the VIS X-Shooter. Here we exclude any telluric contaminated wavelength window. Black lines show the observations and in red we show the best-fitting model spectra. These models use the final abundances obtained as part of this work.}
\label{appfig3}
\end{figure*}

\begin{figure*}
\centering
\includegraphics[angle=270,scale=0.5]{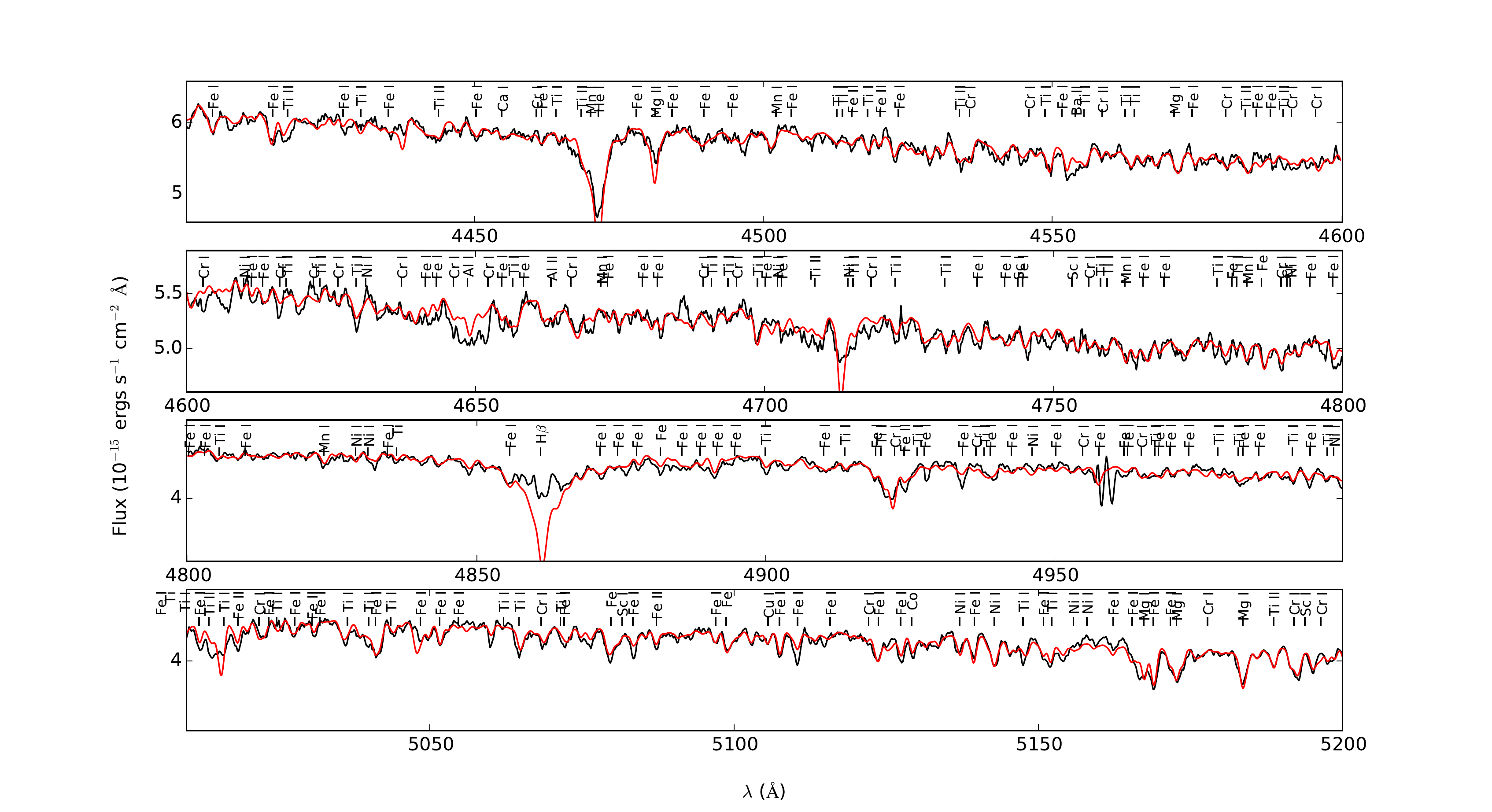}
\caption{Example synthesis fits for NGC1705-1 for different wavelength coverage in the UVB X-Shooter. Black lines show the observations and in red we show the best-fitting model spectra. These models use the final abundances obtained as part of this work. }
\label{appfig1}
\end{figure*}

\begin{figure*}
\centering
\includegraphics[angle=270,scale=0.5]{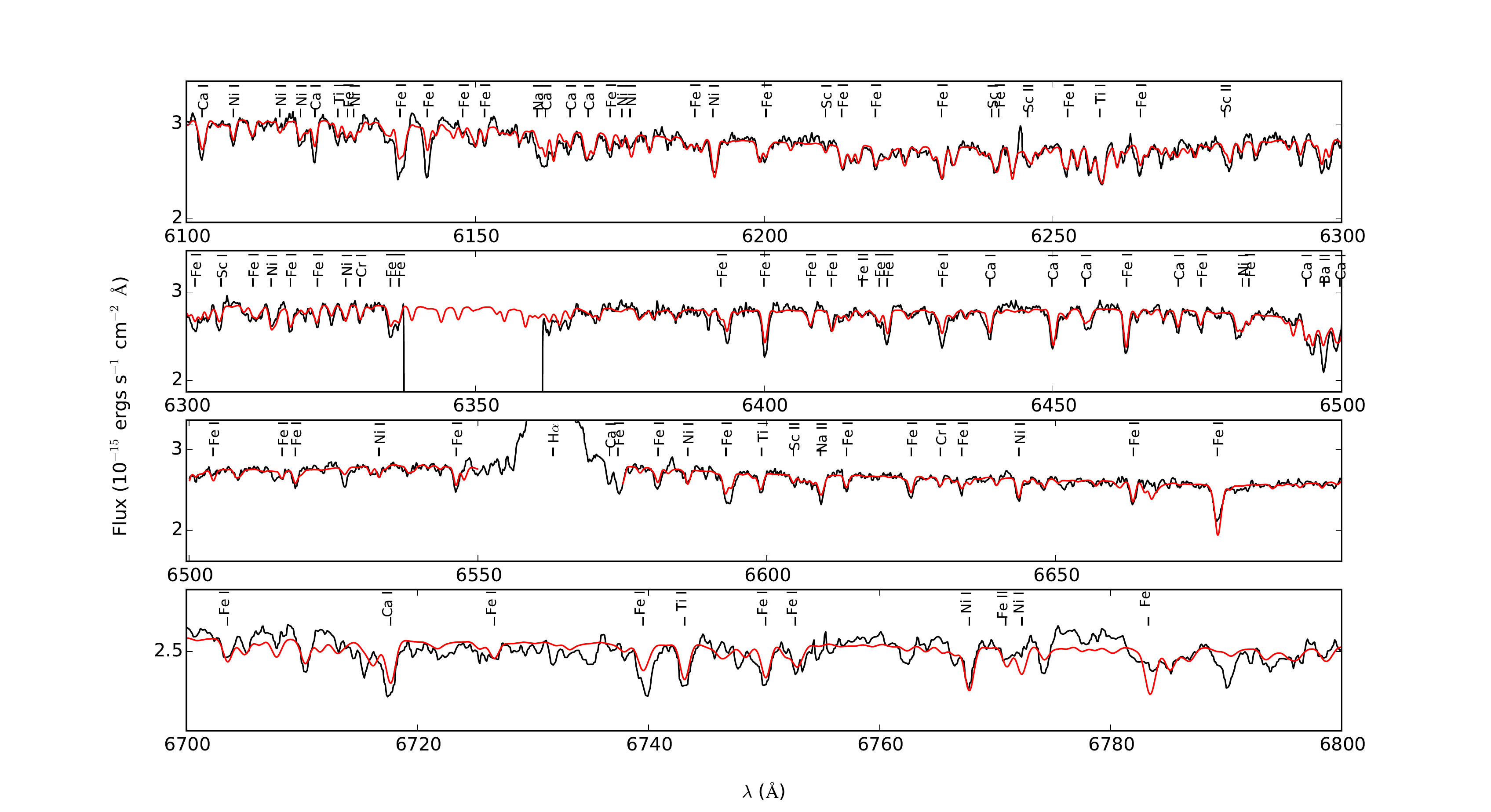}
\caption{Example synthesis fits for NGC1705-1 for different wavelength coverage in the VIS X-Shooter. Here we exclude any telluric contaminated wavelength window. Black lines show the observations and in red we show the best-fitting model spectra. These models use the final abundances obtained as part of this work.}
\label{appfig2}
\end{figure*}

\begin{figure*}
\centering
\includegraphics[angle=270,scale=0.5]{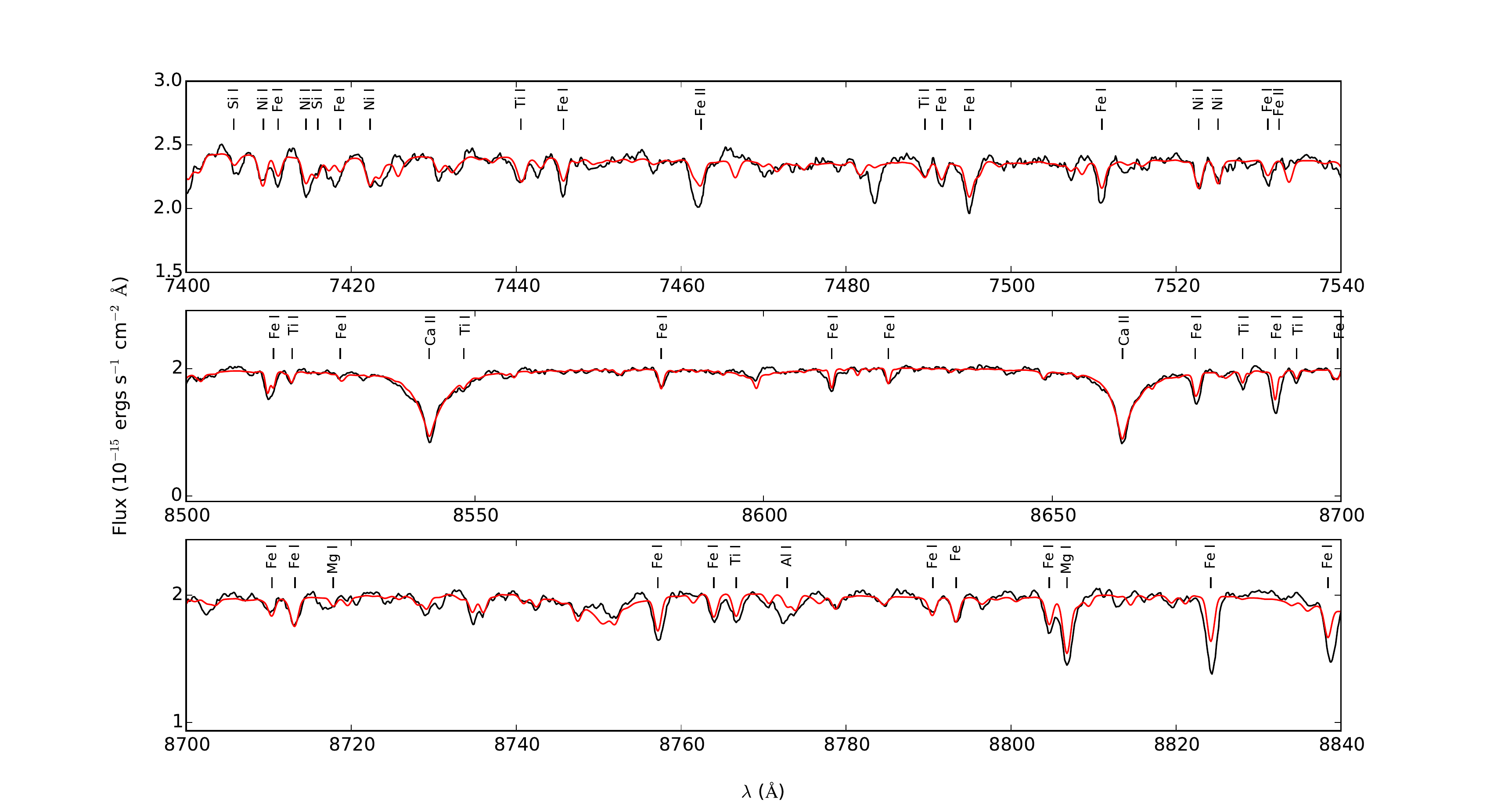}
\caption{Example synthesis fits for NGC1705-1 for different wavelength coverage in the VIS X-Shooter. Here we exclude any telluric contaminated wavelength window. Black lines show the observations and in red we show the best-fitting model spectra. These models use the final abundances obtained as part of this work.}
\label{appfig3}
\end{figure*}

\begin{table*}
\caption{Chemical abundances for NGC1313-379}
\label{table:chem379}
 \centering 
\begin{tabular}{cccc} 
 \hline  \hline \
Element& $Wavelength\, [$\AA$]$&Abundance &Error\\
  \hline\
{[Fe/H]}& 4700.0-4800.0& $-$0.915 & 0.071 \\
& 4900.0-5000.0& $-$1.085 &0.041 \\
& 5000.0-5100.0& $-$1.034&0.051\\
& 6100.0-6300.0&$-$0.914&0.031\\
& 6300.0-6340.0&$-$1.114 & 0.081 \\
& 8500.0-8700.0&$-$0.704 & 0.021\\
& 8700.0-8850.0& $-$0.765 &0.051  \\

{[Mg/Fe]}&5150.0-5200.00&$-$0.310&0.083\\
&8777.0-8832.00&$+$0.299&0.054\\

{[Ca/Fe]}&4445.0-4465.0&$-$0.101&0.251\\
&6100.0-6128.0&$+$ 0.299 &0.122\\
&6430.0-6454.0 &$+$0.378 &0.120\\
&6459.0-6478.0 &$+$0.060&0.193\\
&8480.0-8586.0& $+$0.200& 0.085\\
&8623.0-8697.0 &$+$0.039& 0.018\\

{[Sc/Fe]}& 6222.0-6244.0& $+$0.350& 0.242\\

{[Ti/Fe]}&4650.0-4718.0& $-$0.231& 0.122\\
&4980.0-5045.0& $-$0.049& 0.073\\
&6584.0-6780.0 &$-$0.039 &0.103\\

{[Cr/Fe]}&4580.0-4640.0& -&-\\
& 4640.0-4675.0& $+$0.379 &0.121\\
&4915.0-4930.0& $+$0.619 &0.192\\
&6600.0-6660.0&$+$0.669&0.252\\

{[Mn/Fe]}&4750.0-4770.0 &$-$0.331& 0.252\\

{[Ni/Fe]}& 4700.0-4720.0& $+$0.150 &0.203\\
& 4825.0-4840.0 &$+$0.340 &0.392\\
&4910.0-4955.0& $+$0.330 &0.144\\
&5075.0-5175.0 &$+$0.039 &0.093\\
&6100.0-6200.0 &$+$0.599 &0.093\\
&7700.0-7800.0 &$+$0.950&0.104\\
 \hline 
 \end{tabular}
\end{table*}

\begin{table*}
\caption{Chemical abundances for NGC1705-1}
\label{table:chem1}
 \centering 
\begin{tabular}{cccc} 
 \hline  \hline \
Element& $Wavelength\, [$\AA$]$&Abundance &Error\\
  \hline\
{[Fe/H]}& 4700.0-4800.0&$-$1.137  &0.039  \\
& 4900.0-5000.0&$-$0.797 & 0.028\\
& 5000.0-5100.0& $-$1.152 & 0.033\\
& 6100.0-6300.0&$-$0.735 &0.021\\
& 6300.0-6340.0& $-$0.717  & 0.049\\
& 8700.0-8850.0& $-$0.555 & 0.021 \\

{[Mg/Fe]}&5150.0-5200.00&$-$0.065&0.027\\
&8777.0-8832.00&$+$0.329&0.015\\

{[Ca/Fe]}&4445.0-4465.0& $-$0.440 & 0.331 \\
&6100.0-6128.0&$+$0.900 &0.033\\
&6430.0-6454.0 &$+$0.948&0.061\\
&6459.0-6478.0 &$+$1.146& 0.067\\
&8480.0-8586.0& $+$0.234&0.012 \\
&8623.0-8697.0 &$+$0.040& 0.015\\

{[Sc/Fe]}& 6222.0-6244.0&$+$0.192 & 0.053\\

{[Ti/Fe]}&4650.0-4718.0&$+$0.326 &0.097\\
&4980.0-5045.0& $+$0.652& 0.045\\
&6584.0-6780.0 &$+$0.267 &0.049\\

{[Cr/Fe]}&4580.0-4640.0& $-$0.439&0.092\\
&4640.0-4675.0& - &-\\
&4915.0-4930.0& - & -\\
&6600.0-6660.0&$+$0.666&0.217\\

{[Mn/Fe]}&4750.0-4770.0 &$-$0.229& 0.242\\
{[Ni/Fe]}& 4700.0-4720.0&$-$0.249 &0.142\\
& 4825.0-4840.0 & - &-\\
&4910.0-4955.0&  $-$0.628&0.112\\
&5075.0-5175.0 &  $-$0.178&0.053\\
&6100.0-6200.0 & $+$0.930&0.053\\
&7700.0-7800.0 &$+$1.900& 0.052\\
 \hline 
 \end{tabular}
\end{table*}

\end{appendix}
\end{document}